\documentclass[twocolumn,prb]{revtex4-1}

\usepackage{graphicx}
\usepackage{subfigure}
\graphicspath{ {images/} }

\usepackage{pdfpages}
\usepackage{pgffor}

\makeatletter
\AtBeginDocument{\let\LS@rot\@undefined}
\makeatother

\begin{document}

\title{Electromechanics of Suspended Spiral Capacitors and Inductors}

\author{
	Sina Khorasani
}

\affiliation{
	School of Electrical Engineering, Sharif University of Technology, P. O. Box 11365-9363, Tehran, Iran\\
\'{E}cole Polytechnique F\'{e}d\'{e}ral de Lausanne (EPFL), CH-1015, Lausanne, Switzerland
}
\email{khorasani@sina.sharif.edu; sina.khorasani@epfl.ch}

\begin{abstract}
	
Most electromechanical devices are in two-dimensional metallic drums under high tensile stress, which causes increased mechanical frequency and quality factor. However, high mechanical frequencies lead to small zero-point displacements $x_{\rm zp}$, which limits the single-photon interaction rate $g_0$. For applications which demand large $g_0$, any design with increased $x_{\rm zp}$ is desirable. It is shown that a patterned drum by spiral shape can resolve this difficulty, which is obtained by a reduction of mechanical frequency while the motion mass is kept almost constant. An order of magnitude increase in $g_0$, and agreement between simulations and interferometric measurements is observed. 
	
\end{abstract}

\maketitle

Various applications of electromechanics covers classical and quantum regimes, such as sensing and cryogenic superconducting circuits. Usually, some force such as radiation pressure, acceleration, or gravitational mass is responsible for deformation of a mechanical moving body, or shifting its resonance frequency $\Omega$. In either case, a nonlinear interaction between the mechanical motion of a parallel plate capacitor and the electric component of an oscillating electromagnetic field is developed. The single-photon interaction rate $g_0$, a quantity having units of frequency, defines the strength of such nonlinear electromechanical interactions \cite{1,2}. 

Typical values of $g_0$ are dependent on the application. For optomechanical devices and phoxonic crystals it is on the order of 10MHz or more, while for molecular optomechanics it could take on extremely high values. However, for superconducting electromechanics with micro-drum capacitors, where the reservoir frequency is a few GHz, $g_0$ can be in the range of $2\pi\times 20{\rm Hz}$ to $2\pi\times 60{\rm Hz}$. 

In principle, any method to enhance $g_0$ is favorable and much useful from a practical point of view. All other major applications including mass and force sensing, also rely on $g_0$, so that larger $g_0$ would directly translate into an increased sensitivity, simply because \cite{2} that single-photon cooperativity $\mathcal{C}_0$ is proportional to $g_0^2$. However, field-enhanced cooperativity $\mathcal{C}=\mathcal{C}_0 \bar{n}_{\rm cav}$ may not significantly change since $\bar{n}_{\rm cav}\propto\Omega$ where $\bar{n}_{\rm cav}$ is the equilibrium cavity occupation, implying the fact that $\mathcal{C}$ is independent of $\Omega$. Therefore, the ultimate theoretical side-band cooling limit will remain unchanged, unless squeezed light \cite{2a} or feedback control \cite{2b} schemes are used.

The motivation here is to propose a cost-effective, simple, and efficient method to enhance $g_0$ for a given fabrication process. The trick is to suspend a spiral electromechanical element, letting it vibrate more freely compared to the constrained devices grown on fixed substrates. Here, one may etch a spiral pattern on a drum capacitor, which is shown to be quite feasible. That could be thought of a rolled cantilever, however cantilevers are one-dimensional (1D) structures, while this is effectively a two-dimensional (2D) element with a sensitivity exceeding that of a 1D cantilever. Hence, the cantilever approximation cannot be used, since it would yield incorrect results.

It is also possible to think of suspended inductors, too. While spiral capacitors need to be fixed at one end, suspended inductors should be fixed at both ends to let current flow. In spiral capacitors electrostatic field is responsible for mechanical deformation, while magnetic field of electrical current causes mechanical deformation and pinching of suspended spiral inductors. A major advantage of using suspended inductors is accessing the second quadrature of the electromagnetic radiation field due to the electrical current, while the first quadrature due to the electrical voltage interacts with spiral capacitors. There is otherwise no known method of accessing both quadratures of microwave radiation in a superconductive circuit so easily and in such a straightforward manner. However, suspended inductors have very small $g_0$, typically ranging from 10mHz to 1Hz, as it has been discussed in the Supplementary Information. While apparently too small, these figures are sufficient to be measurable.

\begin{figure}
	\includegraphics[width=\columnwidth]{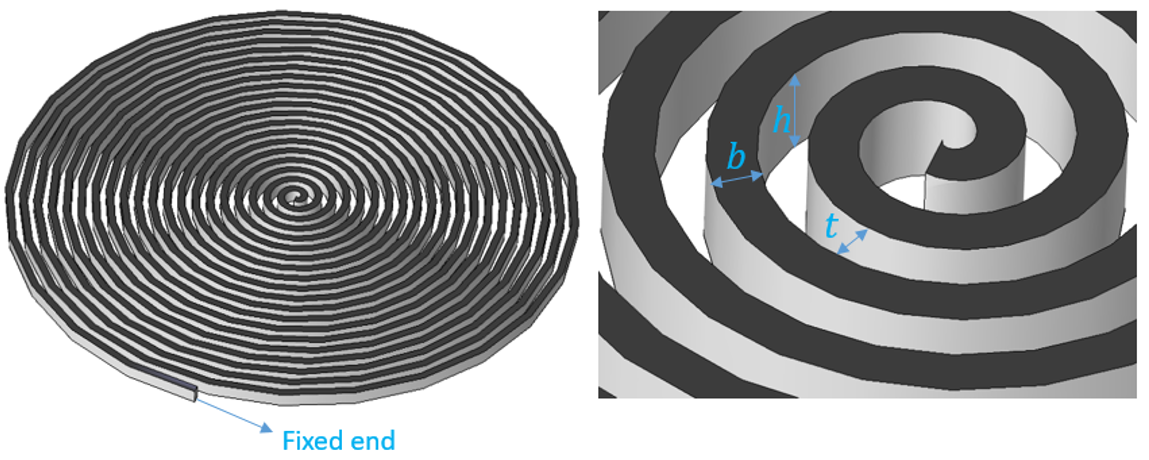}
	\caption{
		Illustration of a uniform spiral. For a spiral capacitor this is the top electrode with one fixed end. For a suspended inductor both ends must be suspended.
		\label{Fig1}
	}
\end{figure}

\begin{figure}
	\includegraphics[width=\columnwidth]{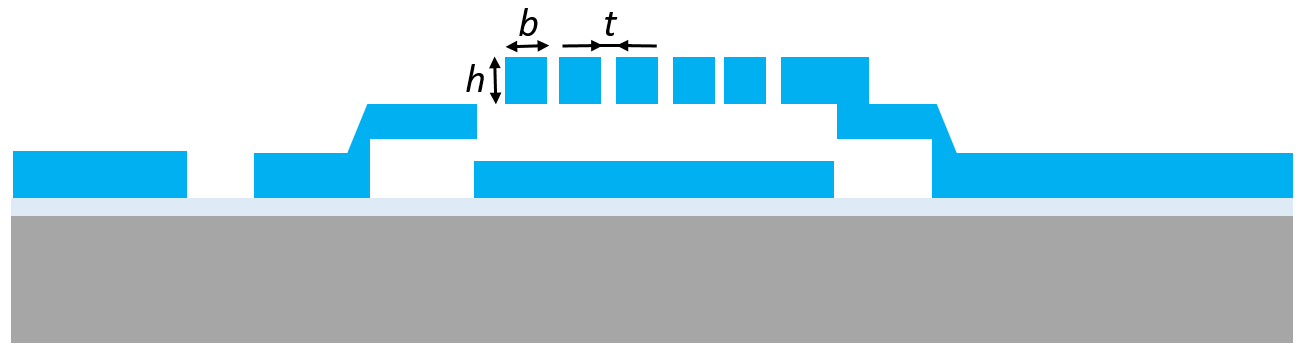}
	\caption{
		Spiral capacitor viewed schematically sideways. Blue color represents Aluminum conducting layers, cyan color is a thin insulating ${\rm SiO_2}$ barrier with the gray Alumina substrate.
		\label{Fig2}
	}
\end{figure}

A spiral structure with uniform spacing and strip width is illustrated in Fig. \ref{Fig1}, with $b$, $h$, and $t$ being respectively the strip width, thickness, and gap as shown in Fig. \ref{Fig2} in sideways. The external radius of a suspended capacitor or inductor at microwave frequencies is typically of the order of $10{\rm \mu m}$ and 1mm, respectively, and the required number of turns $N$ is normally under 20. We find that roughly $g_0\propto \sqrt{N}$ holds, if external parameters are unchanged. Spiral capacitors must be suspended very close to a conducting bottom electrode not exceeding 100nm-150nm. 

The fundamental mechanical mode of a suspended capacitor should clearly be out-of-plane, with one maximum at the center. For this to happen, one should select $b>h$. Strong violation of this condition, however, significantly influences the fundamental mode, making it in-plane, much like the clock spiral springs. This design is obviously favorable for the suspended inductor.

By cutting through a spiral and undercutting of the capacitor the initial tensile stress is suddenly removed, and thus the spiral is expected to contract horizontally after release. This may lead to difficulties in fabrication, and it is at first not quite obvious that this structure actually can be made. 

The other issue could be Al grain size which puts a practical limit on the achievable minimum $b$. Atomic Layer Deposition (ALD) of Al has been reported \cite{3}, but is not customary, and the deposited Al using evaporation is not crystalline. Replacement superconducting metals which could be grown in crystalline form are not known. This implies that for the moment being, and while not having access to a crystalline growth of Aluminum (through ALD, MBE, etc.), one would need to find a practical solution to demonstrate the feasibility of process. This puts severe restrictions on the fabrication. 

At first, it is rather hard to imagine that the spiral capacitors could survive the undercut. An evaluation of the idea would suggest that the undercut and released spiral would collapse, buckle, break because of Van der Waals attraction, or at least significantly deform out of plane because of thermal coefficient mismatch. It could be so fragile that would break while carriage. Neither of these happened, contrary to the normal expectations, and we can show here that the spiral capacitor with moderate number of turns, can be successfully fabricated and suspended. We have furthermore measured the mechanical response and observed complete agreement to the design. 

While we cannot satisfactorily explain yet why the structure survives the fabrication and undercut, possible explanations are first that Focused Ion Beam (FIB) process could infuse and crystallize Al grains, making the grown layer effectively in terms of mechanical properties much like a single-crystal. Secondly, the suspended Al is bounded to vacuum from both top and bottom sides after undercut and is too thin (100nm) to develop any significant residual stress gradient during growth. Hence, it does not buckle up or down similar to what always happens to the multilayer or very thick cantilevers, which are highly deformed after release and undercut. 

\begin{figure}
	\includegraphics[width=4cm]{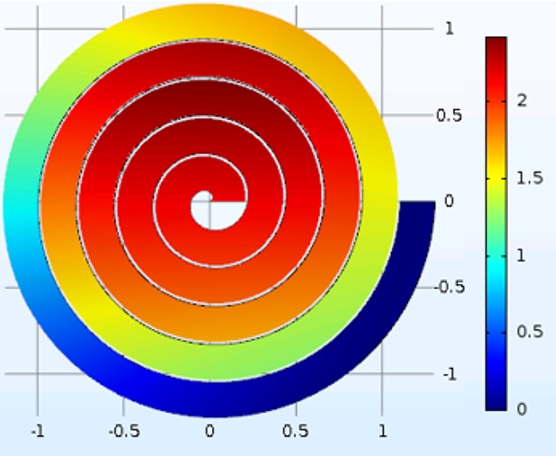}
	\caption{
		COMSOL simulation of spiral with $N=5$. The fundamental frequency is found to be $f=20.96{\rm kHz}$. Colors correspond to the polar out-of-plane deformation. The fundamental mechanical mode is out-of-plane, which makes the structure appropriate to function as a spiral capacitor.\label{Fig3}
	}
\end{figure}

Table \ref{Table1} summarizes various spiral geometries on the same structure. Calculations are done using the polar deformation profile $\Delta\rho(\theta)$ fed from COMSOL, as discussed in the supplementary material. The first row corresponds to the simple membrane of the unpatterned micro-drum capacitor, which exhibits a much larger mechanical frequency due to the residual tensile stress. The second row corresponds to what is fabricated with $N=5$, whose COMSOL simulation is illustrated in Fig. \ref{Fig3}. . By carving only 5 and 10 turns, $g_0$ increases respectively 7- and 12-fold, showing a rough dependence $g_0\propto\sqrt{N}$. For inductor simulations on COMSOL, both ends should be fixed and the first in-plane displacement mode is investigated. By choosing sufficient thickness, the fundamental mode becomes in-plane polarized. For spiral capacitors, a small $N$ is quite sufficient to obtain large enhancement of $g_0$. The deformation profile can be also estimated theoretically within the thin-wire approximation as detailed in the Supplementary Information. But that would mostly cause underestimation of $g_0$. 

\begin{table}
	\begin{center}
		\begin{tabular}{c c c c c c c c}
			\hline\hline
			$b$ & $h$ & $t$ & $d$ & $L_0 ({\rm nH})$ & $f ({\rm kHz})$ & $N$ & $g_0 ({\rm Hz})$ \\
			\hline
			$-$ & 100 & $-$ & 100 & 70 & $6.2\times10^3$ & 0 & $2\pi\times60$ \\
			2000 & 100 & 200 & 100 & 70 & 20.96 & 5 & $2\pi\times418$ \\
			1000 & 100 & 200 & 100 & 70 & 10.5 & 10 & $2\pi\times701$ \\
			1000 & 100 & 100 & 100 & 70 & 1.63 & 20 & $2\pi\times941$ \\
			\hline\hline
		\end{tabular}
	\end{center}
	\caption{
		Typical $g_0$ for various spiral capacitor configurations \cite{4}. The first row corresponds to the unpatterned structure under tensile stress. The second row corresponds to what has been fabricated. Dimensions of $b$, $h$, $t$, and $d$ are given in nm.
		\label{Table1}
	}
\end{table}

There is no parasitic resistance in the lumped equivalent circuit of the spiral capacitor, but for non-superconducting states,  it can be easily derived by simple geometrical considerations. The parasitic inductance \cite{5} at microwave frequencies of interest is also not a matter of concern, since the typical wavelength is orders of magnitude larger than the spiral diameter. Hence, the spiral is essentially so small that it remains equipotential everywhere and any parasitic inductor can be neglected.

We did not try fabrication of suspended inductor, despite easier fabrication due to much larger size. For spiral capacitors, the motion mass $m$ is roughly $2/3$ of the total mass. One can here estimate $\Omega$ from a 1D cantilever equivalent \cite{6,7,8,9} with the identical $b$, $h$, and curve length. However, the result is normally off the correct value by two orders of magnitude, or even more. Hence, the deformation profiles and $\Omega$ must be found numerically for good accuracy, as it has been extensively discussed in the supplementary material. Now, $x_{\rm zp}=\sqrt{\hbar/2m\Omega}$ can be found, which yields $g_0=x_{\rm zp}(\partial\omega/\partial x)$. 

\begin{figure}
	\includegraphics[width=\columnwidth]{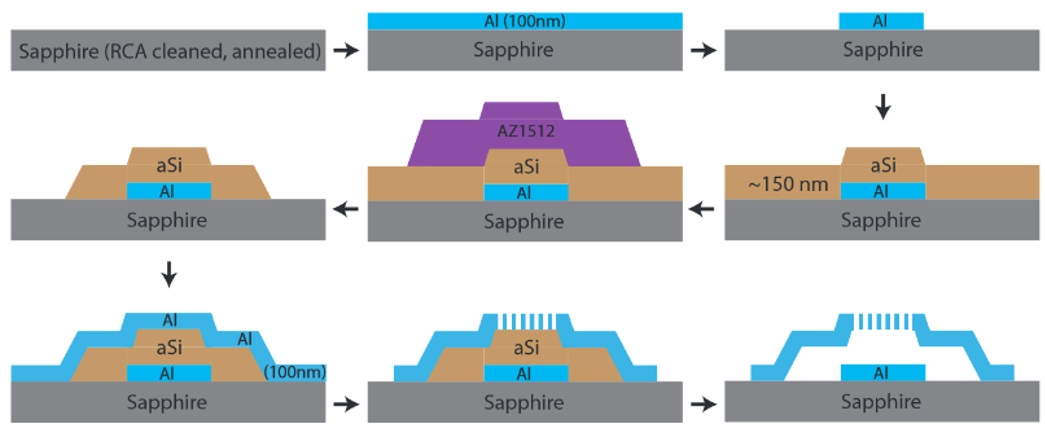}
	\caption{
		Fabrication process flow is based on the drum capacitor of an earlier study \cite{10}, except that a spiral pattern on the top electrode is produced using FIB.
		\label{Fig4}
	}
\end{figure}

It is possible to redesign and remake all masks by EBL, so that the additional FIB could be removed. Should masks need redesign to be fabricated with EBL, then there exist macros such as the one in Supplementary Information which could easily put a spiral to the mask design with desired shape parameters. However, UV lithography would not be possible anymore and the compatibility of E-Beam photoresists with the present process has yet to be investigated. The accuracy of UV lithography is good enough to support fabrication of suspended inductors if necessary. However, $h$ probably needs to be much more than 100nm to provide sufficient mechanical strength under pinching force of magnetic field. 

We also are not completely unsure of the irrelevance of FIB to successful release, and these have yet to be investigated in a much deeper study. Therefore, unless a rigorous process has to be developed from the scratch, probably the most straightforward way to fabricate a spiral capacitor is to use the already available micro-drum capacitors \cite{10,11} before undercut and release, take them to the FIB, do the patterning, and then carry out the structure release and undercut at last. The fabrication process flow used for this device is presented in Fig. \ref{Fig4}.

\begin{figure}[h!]
	\centering
	\subfigure{
		\includegraphics[width=3cm]{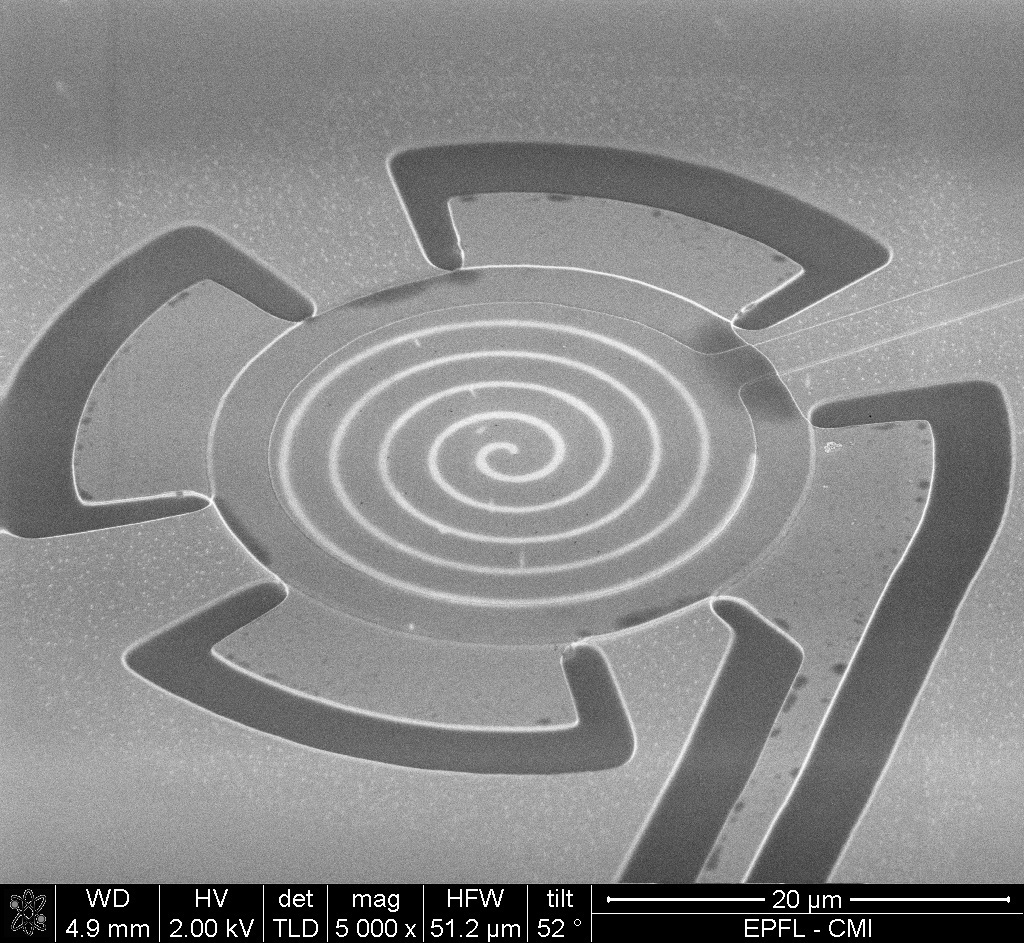}
	} 	
	\\(a)\\
	\subfigure{
		\includegraphics[width=3cm]{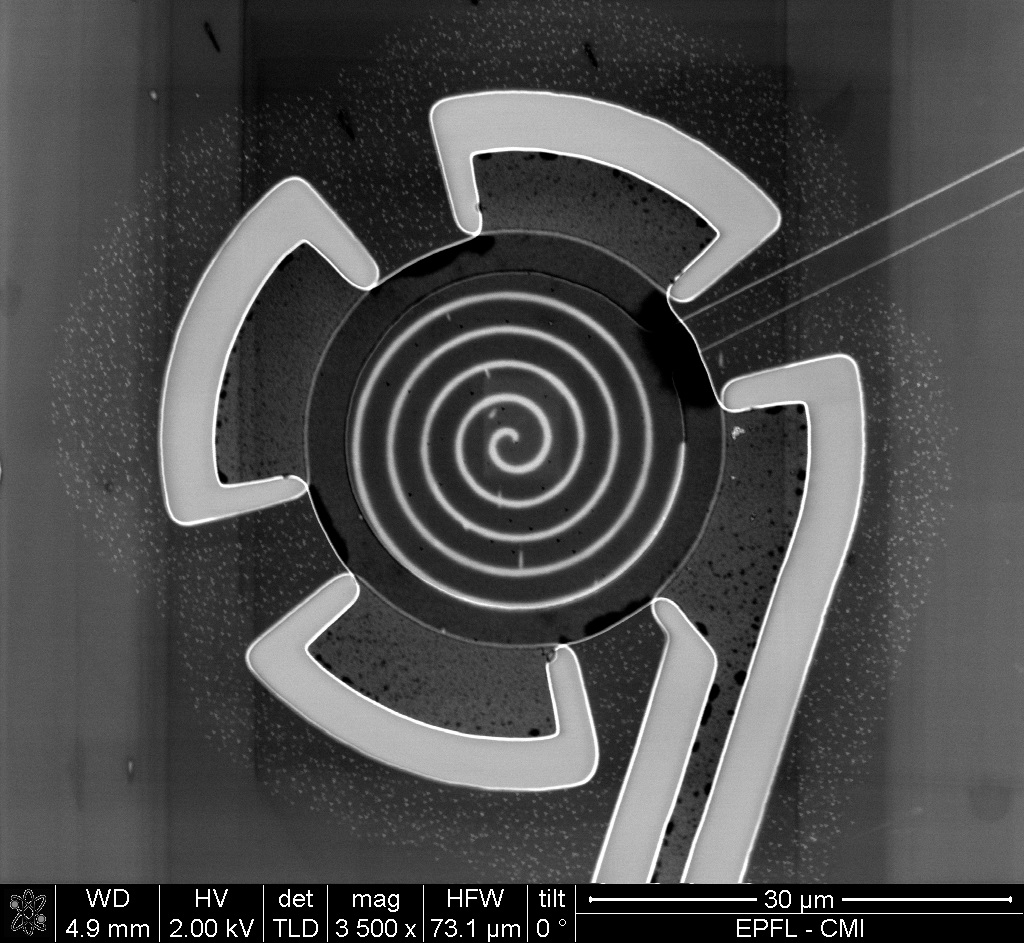}
	}
	\\(b)\\
	\subfigure{
		\includegraphics[width=3cm]{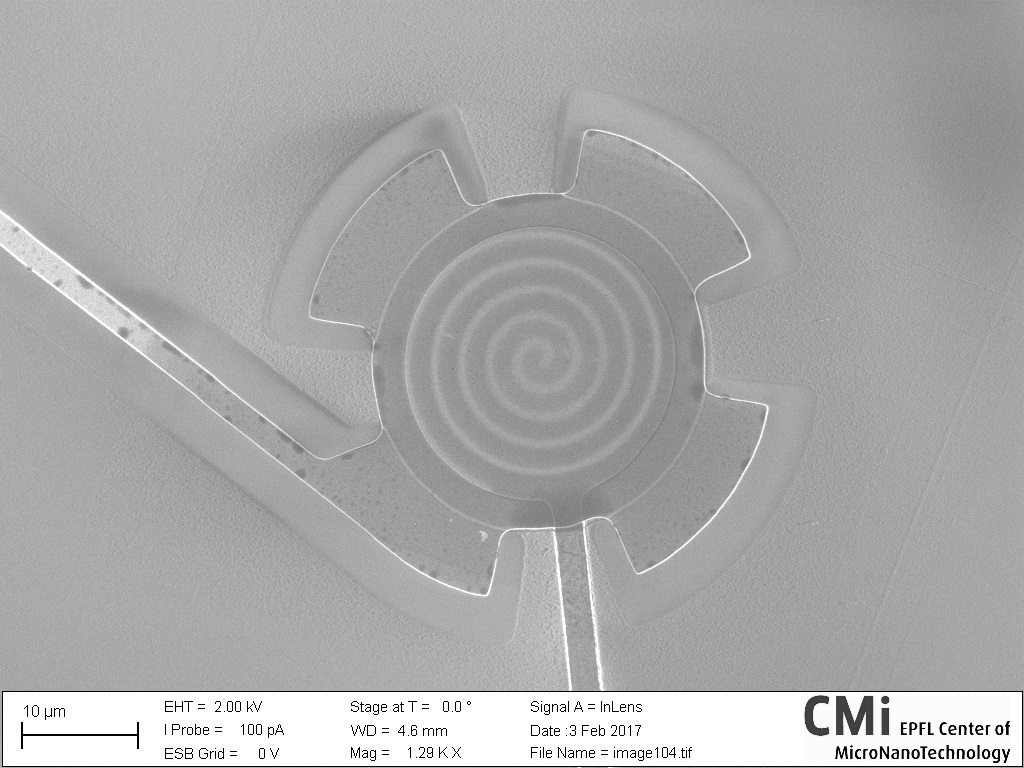}
	}
	\\(c)
	\caption{
		(a) SEM after FIB patterning of the second sample; (b) Ion-beam photo after FIB of second sample; (c) Normal SEM after ${\rm XeF}_2$ undercut.
		\label{Fig5}
	}
\end{figure}

The FIB machine provides both SEM and Ion-beam images at once, while patterning. Figures \ref{Fig5}a and \ref{Fig5}b respectively illustrate those images of the fabricated spiral. After performing the FIB, the sample was taken to the undercutting process with the gaseous ${\rm XeF}_2$ etch. It can be seen in the SEM photo in Fig. \ref{Fig5}c that it has survived the undercut very well. It was possible to carry it around afterwards quite safely to the SEM zone and take a few additional SEM photos at normal and oblique incidences. It should be added that since $t<200{\rm nm}$, the spiral is totally invisible under optical microscope. 

\begin{figure}[h!]
	\centering
	\subfigure{
		\includegraphics[width=\columnwidth]{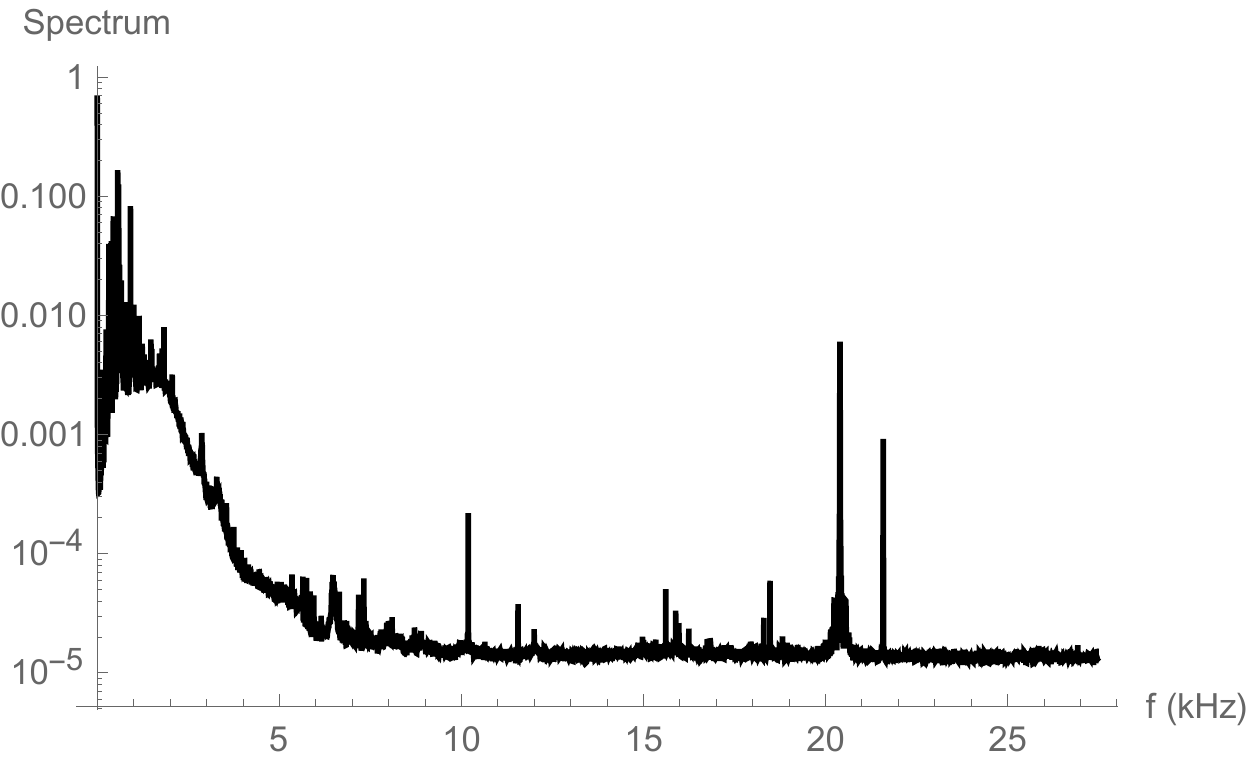}
	}
	\\(a)\\
	\subfigure{
		\includegraphics[width=\columnwidth]{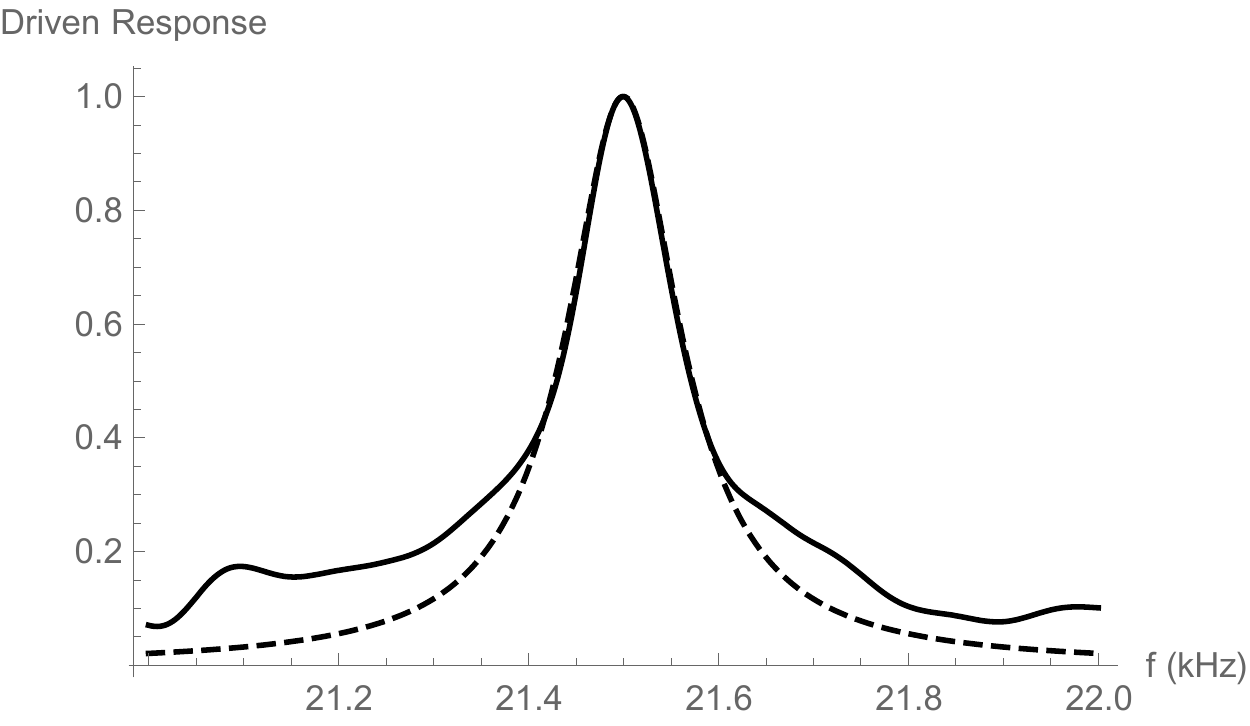}
	}
	\\(b)\\
	\subfigure{
		\includegraphics[width=\columnwidth]{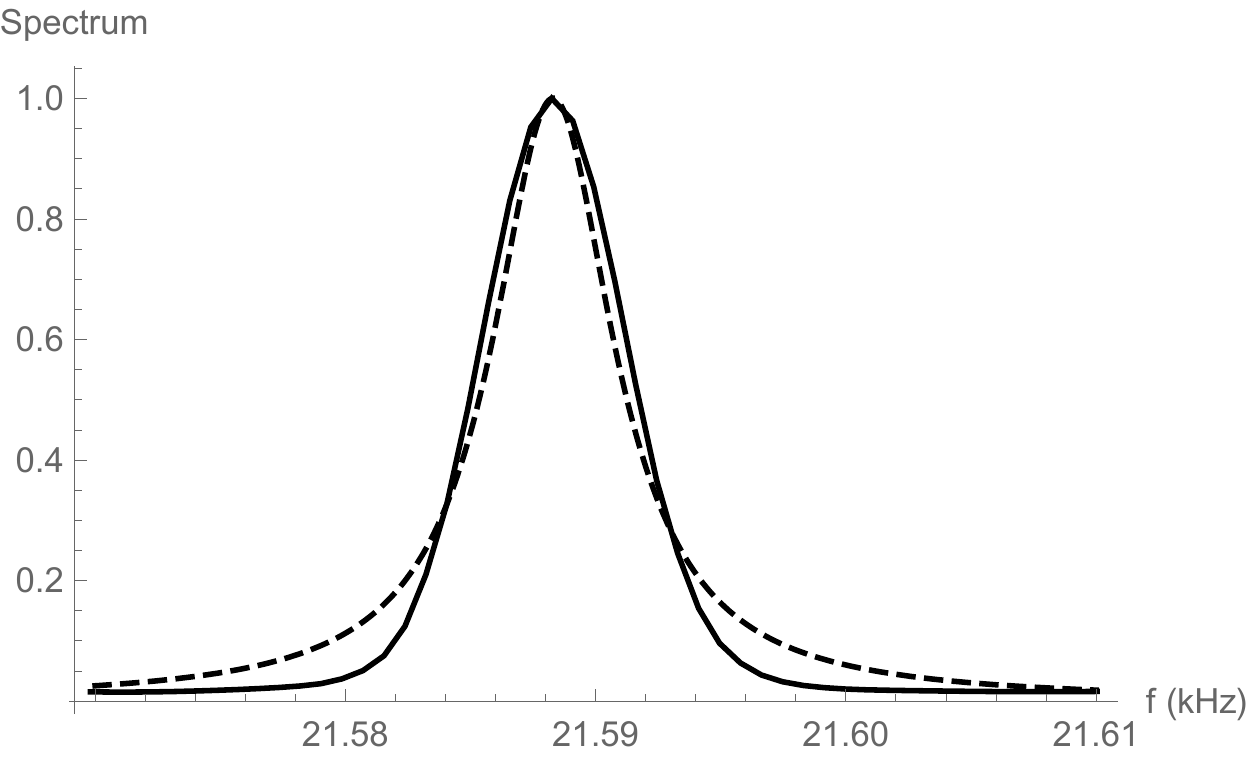}
	}
	\\(c)
	\caption{
		Measurements of mechanical response using optical interferometry under high vacuum and at room temperature: (a) Natural response; (b) Driven response around the resonance with $Q=148$ and $f=21.5{\rm kHz}$; (c) Measured response with $Q=3.6\times 10^3$ and $f=21.6{\rm kHz}$.
		\label{Fig6}
	}
\end{figure}

The mechanical response of the fabricated spiral capacitor was tested at room-temperature while the sample was placed in a high-vacuum chamber with transparent quartz window, mounted on an isolated optical table. The reflection of a continuous red laser from spiral surface at normal incidence was fed into an interferometric setup, allowing precise observation of the mechanical movements. The sample holder is mounted on a piezo-electric actuator which can be biased and excited by a sinusoidal frequency. 

Measurements were done under two modes: (i) the natural response due to thermal fluctuations with no external mechanical excitation, and (ii) the driven or forced response under sinusoidal excitation of the piezo-electric actuator with tunable frequency drive.

The driven measurement is needed to identify the right resonance peak, since the natural response of the mechanical structure is exhibits many spurious modes resulting from substrate and sample mount, along with presence of $1/f$ noise. This has been illustrated in Fig. \ref{Fig6}a. The driven measurement let us identify a clear and unmistakable resonance, as shown in Fig. \ref{Fig6}b. But the resulting $\Omega$ and quality factor $Q$ are not accurate, because of the large mechanical power delivered to the sample causing shifts in original values. Once the actual resonance is found by drive, the accurate $\Omega$ and $Q$ can be clearly derived from the natural response.

The results of measurements across the fundamental resonance are displayed in Figs. \ref{Fig6}b,c. Quite obviously, the two measurement modes are not exactly the same, and both $\Omega=2\pi\times f$ and $Q$ are different. While the driven response yields a roughly $Q=148$ and $f=21.5{\rm kHz}$, the natural response exhibits roughly $Q=3.6\times10^3$ and $f=21.6{\rm kHz}$ at room temperature. As discussed in the above, we here pick the latter values. 

Interestingly, COMSOL simulations of the natural or free response predicts $f_0=20.96{\rm kHz}$, which is in very good agreement with both measurement modes. This also very well confirms the accuracy of numerical simulations as well as successful fabrication and levitation of the spiral despite very narrow gap from substrate.

It is extremely difficult to theoretically estimate $Q$ of spirals, and perhaps the straightforward way is to fabricate them and measure their response. Nevertheless, the known mechanisms which limit $Q$ may be quite different, including loss due to finite viscosity of the chamber ambient pressure, phonon tunneling \cite{12} from coupling of the spiral tail to the mount, friction loss between Al grain boundaries, finite electrical conductivity of the non-superconducting Al coupled to the mechanical motion \cite{13,14,15} at the higher temperatures, and ultimately the quantum electrodynamical friction of vacuum \cite{16,17}.

It is a well-known, yet not theoretically explained, experimental fact that measurements on superconducting mechanical oscillators below the critical transition temperature usually causes a typical four- to ten-fold increase in $Q$. Hence, it could be expected that at temperatures  which Al superconducts, $Q$ could still increase to much higher values. Unfortunately, the present experimental setup of our interferometric measurement is not cryogenically cooled and maintains only the room-temperature, disallowing further investigation of this fact. But similar observations have been made on amorphous silica \cite{18}, which reveal a significant increase in $Q$ up to three orders of magnitude at cryogenic temperatures.

Presented designs are only for uniform spirals. Non-uniform spirals could possibly still lead to improved $g_0$ without degrading noise performance, which has been shown to be true for inductors in the Supplementary Information. Finally, suspended inductors might be fabricated along with spiral capacitors, to permit access to both quadratures of the electromagnetic field. The capacitor should be remade, put in an LC circuit, and tested to make sure that it is not short circuited inside. Its noise performance should be carefully investigated, since lower $\Omega$ implies larger phonon occupation number.

In conclusion, a method to enhance $g_0$ was presented for electromechanical quantum superconducting circuits as well as sensing applications. A detailed theoretical model and numerical simulation was developed. Effects of various parameters were studied and it was demonstrated that FIB could provide an easy means for rapid prototyping, without any need to redesign or optimize the earlier fabrication masks and steps. It was shown that the spirals can survive ${\rm XeF}_2$ undercut. The results of this study opens up new possibilities and applications in sensing, electromechanics, quantum circuits \cite{19}, and other sorts of electromechanical systems.

\section*{Supplementary Material}
See supplementary material for details of the developed thin-wire formalism, design of spiral inductors, spiral capacitors, numerical COMSOL simulations, and fabrication process flow.

\begin{acknowledgments}
	Discussions with Prof. Guillermo Villanueva, Dr. Cyrille Hibert, Dr. Philippe Langlet, Dr. Christophe Galland, Dr. Alexey Feofanov, and Mr. Amir Hossein Ghadimi is appreciated. Initial unpatterned micro-drum capacitor was provided by Daniel T\'{o}th. Fabrication was done at the Center for Microtechnology (CMi) of EPFL, with the help of Ryan Schilling, Cl\'{e}ment Javerzac-Galy, Dr. Joffrey Pernollet, and Ms. Nahid Hosseini. Interferometric characterization was done by Dr. Nils Johan Engelsen. This work has been supported by Laboratory of Photonics and Quantum Measurements (LPQM) at EPFL and Research Deputy of Sharif University of Technology.
\end{acknowledgments}



\section*{Supplementary material}

# Supplementary Material
# Electromechanics of Suspended Spiral Capacitors and Inductors

Sina Khorasani

## 1. DESIGN

This section presents a theoretical model for the suspended inductor and capacitor, which ultimately forms an integral equation in the thin wire limit. This can be evaluated numerically to find the solution. Although comparison to COMSOL simulations reveal that the thin wire limit might be good only within the order-of-magnitude estimation, however, the usefulness of the proposed model to provide insight to the behavior of the structure could not be ignored. We first present the model and then numerical case studies are discussed.

### 1.1. Derivation of Strained Shape

*1.1.1. Suspended Spiral Inductor*

Calculations for the strain of a spiral inductor due to the current is conveniently done by switching from polar coordinates to complex plane. If the spiral is defined by the polar function $\rho(\theta)$, then the corresponding complex number in the 2D complex plane is $\bar{\rho}(\theta) = \rho(\theta)\angle\theta = \rho(\theta)e^{i\theta}$. We may switch back from the complex plane to the normal 2D plane anytime later easily by means of geometric one-to-one correspondence between these two coordinate systems. We define $\bar{\rho}(\theta)$ as for the unstrained inductor and $\bar{\varrho}(\theta)$ for the strained one. Hence, the displacement is defined as

$$\Delta\bar{\rho}(\theta) = \bar{\varrho}(\theta) - \bar{\rho}(\theta) = [\varrho(\theta) - \rho(\theta)]e^{i\theta} = \Delta\rho(\theta)e^{i\theta} \tag{S1}$$

We also define

$$\bar{u}(\theta) = \bar{\rho}'(\theta) = [\rho'(\theta) + i\rho(\theta)]e^{i\theta} \tag{S2}$$

Hence, we obtain

$$\bar{\varrho}(\theta) = \bar{\rho}(\theta) + \Delta\bar{\rho}(\theta) = \int_0^\theta \bar{u}(\beta) + \int_0^\beta \Delta\bar{u}(\gamma)\,d\gamma d\beta \tag{S3}$$

For a curved segment with the length $\Delta l$, the deformation under uniform force $F$ exerted by the magnetic field is given by

$$\Delta\phi = \frac{F}{6EM}\Delta l^2 \tag{S4}$$

where $\Delta l = \rho(\theta)d\theta$. We may also note that

$$\Delta\bar{u} = \bar{u}e^{i\Delta\phi} - \bar{u} \approx i\bar{u}\Delta\phi \tag{S5}$$

Therefore, we obtain

$$\Delta\bar{u} = i\frac{F\bar{\rho}'(\theta)}{6EM}\bar{\rho}^2(\theta)\Delta\theta^2 \tag{S6}$$

Twice integration of this expression gives the appropriate equation for displacement as

$$\Delta\bar{\rho}(\theta) = \int_0^\theta\int_0^\beta \Delta\bar{u}(\gamma)\,d\gamma d\beta = \frac{i}{6EM}\int_0^\theta\int_0^\beta F(\gamma)\bar{\rho}'(\gamma)\bar{\rho}^2(\gamma)\,d\gamma d\beta \tag{S7}$$

or

$$\bar{\varrho}(\theta) = \bar{\rho}(\theta) + \frac{i}{6EM} \int_0^\theta \int_0^\beta F(\gamma)\bar{\rho}'(\gamma)\bar{\rho}^2(\gamma)\, d\gamma d\beta$$

(S8)

Differentiating both sides once gives the differential equation

$$\bar{\varrho}'(\theta) = \bar{\rho}'(\theta) + \frac{i}{6EM} \int_0^\theta F(\gamma)\bar{\rho}'(\gamma)\bar{\rho}^2(\gamma)\, d\gamma$$

(S9)

Separation of real and imaginary parts gives out nonlinear integro-differential equations of order two and one, respectively. After some algebra, the imaginary part can be recovered and reads

$$\varrho'(\theta) + A \int_0^\theta \rho'(\gamma)\rho^3(\gamma) \int_0^\Theta \frac{\Re\{[\bar{\varrho}(\beta) - \bar{\varrho}(\gamma)]e^{-j\beta}\}}{|\bar{\varrho}(\beta) - \bar{\varrho}(\gamma)|^3}\varrho(\beta) d\beta\, d\gamma = \rho'(\theta)$$

(S10)

where

$$A = \frac{\mu_0 I^2}{48\pi EM}$$

(S11)

and by using the Biot-Savart law of Magnetic force we have plugged-in

$$\frac{dK(\theta)}{dl} = \frac{F(\theta)d\theta}{\varrho(\theta)d\theta} = \frac{\mu_0 I^2}{4\pi} \int_0^\Theta \frac{\Re\{[\bar{\varrho}(\beta) - \bar{\varrho}(\theta)]e^{-j\beta}\}}{|\bar{\varrho}(\beta) - \bar{\varrho}(\theta)|^3}|\bar{\varrho}(\beta)| d\beta$$

(S12)

Furthermore, the difference between $\bar{\varrho}(\theta)$ and $\bar{\rho}(\theta)$ can be ignored to reach

$$F(\theta) \cong \frac{\mu_0 I^2}{4\pi} \rho(\theta) \int_0^\Theta \rho(\beta) \frac{\Re\{[\bar{\rho}(\beta) - \bar{\rho}(\theta)]e^{-i\beta}\}}{|\bar{\rho}(\beta) - \bar{\rho}(\theta)|^3} d\beta$$

(S13)

which is to avoid unnecessary complication in calculations. Hence, we have

$$\Delta\rho(\theta) \cong -A \int_0^\theta \rho'(\beta)\rho^3(\beta) \int_0^\Theta \frac{\rho(\gamma)\Re\{[\bar{\rho}(\gamma) - \bar{\rho}(\beta)]e^{-i\gamma}\}}{|\bar{\rho}(\gamma) - \bar{\rho}(\beta)|^3} d\gamma\, d\beta$$

(S14)

that is the used form in the above.

In practice, evaluation of the inner integral is tough, and needs extra care because of the logarithmic divergence and highly oscillatory integrand. It is possible to overcome this obstacle by introducing a finite positive parameter $\epsilon$ as

$$\Delta\rho(\theta) \cong -A \int_0^\theta \rho'(\beta)\rho^3(\beta) \int_0^\Theta \frac{\rho(\gamma)\Re\{[\bar{\rho}(\gamma) - \bar{\rho}(\beta)]e^{-i\gamma}\}}{[\epsilon + |\bar{\rho}(\gamma) - \bar{\rho}(\beta)|]^3} d\gamma\, d\beta$$

(S15)

Now, letting $\epsilon$ approach zero makes it exact. Choosing $\epsilon = 0.1a$ works fine, where $a$ is a constant of unit length, used in defining the shape of the spiral inductor. It has been observed that smaller values lead to problems in stability of the integration technique.

*1.1.2. Parallel-plate Capacitor*

For a parallel-plate capacitor with a flat and suspended spiral electrode, the force per unit length is almost constant and given by

$$\frac{dK(\theta)}{dl} = \epsilon_0 \frac{b}{[d - \Delta\rho(\theta)]^2} V^2$$

$$\Delta\rho(\theta) \cong d^2 B \int_0^\theta \rho'(\beta)\rho^3(\beta) \frac{1}{[d - \Delta\rho(\beta)]^2} d\beta$$

$$B = \frac{\epsilon_0 b V^2}{12\pi EM}$$

(S16)

This greatly simplifies if we make the assumption that $d \gg \Delta\rho(\theta)$. Then we have

$$\frac{dK(\theta)}{dl} \cong \epsilon_0 \frac{b}{d^2} V^2$$

$$\Delta\rho(\theta) \cong B \int_0^\theta \rho'(\beta)\rho^3(\beta) \, d\beta$$

$$B = \frac{\epsilon_0 b V^2}{12\pi EM d^2}$$

(S17)

We further note that the displacement $\Delta\rho(\theta)$ for the case of inductor is in-plane and corresponds to a displacement in the $x - y$ plane, while for the capacitor is off-plane and corresponds to a displacement along $z$ −direction.

*1.1.3. Effective Motion Mass*

If we simply assume that the total displacement is nearly linearly increasing along the length of the spiral, then the motion mass corresponding to the fundamental harmonic would be simply two-third of the total wire mass as shown below. If $\tau$ denotes the mass density of Aluminum, then we have the relationship for the motion mass as

$$m = bh\tau \frac{\int_0^\Theta \rho(\theta)|\Delta\rho(\theta)|d\theta}{\frac{1}{\Theta}\int_0^\Theta |\Delta\rho(\theta)|d\theta}$$

(S18)

Now, assumption of a linear variation for the displacement roughly as $\Delta\rho(\theta) \sim k(\Theta - \theta)$ where $k$ is some constant, and setting $\rho(\theta) = a\theta$ we get

$$m = bh\tau \frac{\frac{1}{6}ak\Theta^3}{\frac{1}{2}k\Theta} = bh\tau \frac{a\Theta^2}{3}$$

(S19)

But the total mass is

$$M = bh\tau \int_0^\Theta \rho(\theta)d\theta = bh\tau \frac{a\Theta^2}{2}$$

(S20)

Hence, the motion mass is

$$m = \frac{2}{3}M$$

(S21)

**1.2. Suspended Inductor**

An inductor with 12.5 turns is assumed, made out of superconducting Aluminum wire with thickness $h$ and width $b$ of 50nm. The spacing between unperturbed spiral branches is $10\mu m$, so that the spiral inductor fits in a circle area of less than 0.8mm in radius. The shape of inductor may be defined in the polar coordinates as the function $\rho(\theta)$. The return path of the current is momentarily not included in the calculation, and the inductor is coupled to a tank capacitor of 50fF. With no current flowing, the inductance value is 86.5nH, which gives us a resonant frequency value of 2.4GHz. The inductor with no current flowing looks like the following

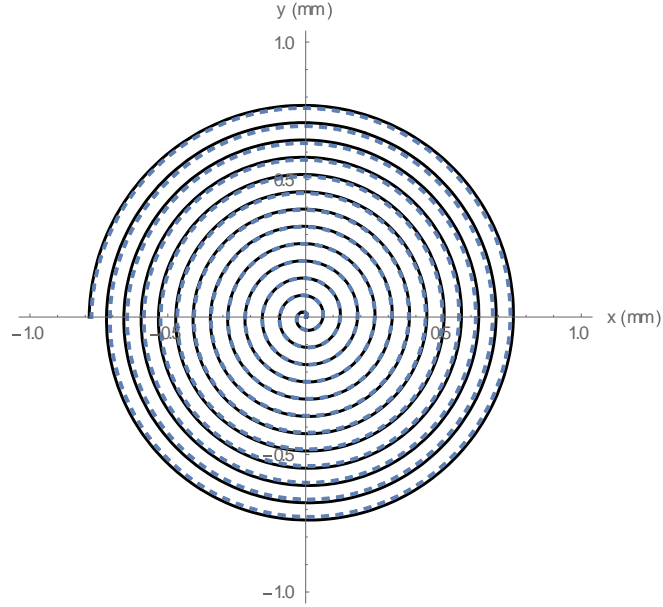

**Fig. S1.1.** Spiral inductor in original (solid black) and after in-plane deformation (dashed blue) due to the pinching force of the magnetic field caused by spiraling current. Both inward and outward currents cause in-plane contraction, so that the pinching force is proportional to the squared current $I^2$.

It is possible to construct an integral representation of the displacement in polar coordinates, which was found in Section 1.1.1, to be given by

$$\Delta\rho(\theta) = -\frac{\mu_0 I^2}{48\pi EM}\int_0^\theta \rho'(\beta)\rho^3(\beta)\int_0^\Theta \frac{\rho(\gamma)\Re\{[\bar{\rho}(\gamma)-\bar{\rho}(\beta)]e^{-i\gamma}\}}{|\bar{\rho}(\gamma)-\bar{\rho}(\beta)|^3}d\gamma\, d\beta$$

(S22)

where $\bar{\rho}(\gamma) = \rho(\gamma)e^{i\gamma}$, $I$ is the electric current, $E$ is the Elastic modulus of Aluminum, and $M = \frac{1}{12}hb(h^2+b^2)$ is the area moment of inertia where $h$ and $b$ are respectively the thickness and width of the Aluminum strip. Furthermore, $\Theta$ is the maximum polar angle at which the spiral terminates.

When the current flows, the magnetic field stress makes the inductor to contract, and as a result the inductance also decreases. For an exaggerated counter-clockwise current of $I = 200 nA$, the displaced inductor is shown as the blue dashed curve (somewhat exaggerated for illustration purposes). The displacement at the outer end of the spiral could be plotted against the current and change in the self-inductance shown below

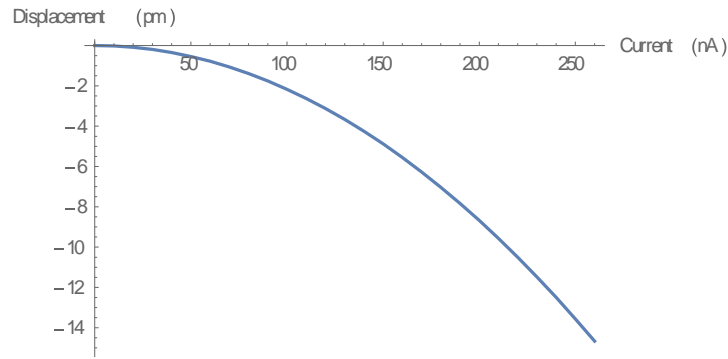

**Fig. S1.2.** Typical in-plane displacement of a suspended spiral thin wire inductor versus current.

The inductance can be approximated from the relationship [5]

$$L \cong \frac{\mu_0}{4\pi}\int_0^\Theta\int_0^\Theta \frac{\Re\{\bar{\rho}'(\gamma)\bar{\rho}'^*(\beta)\}}{|\bar{\rho}(\gamma)-\bar{\rho}(\beta)|}d\gamma d\beta + \frac{\mu_0}{4\pi}Y\int_0^\Theta |\bar{\rho}(\gamma)|d\gamma$$

(S23)

4

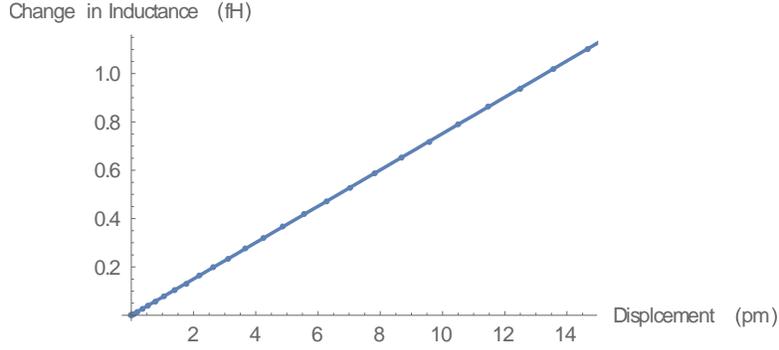

**Fig. S1.3.** Expected change of inductance due to the displacement. Since the displacement is too small compared to the overall size of the inductor, the behavior is linear to a high accuracy.

where $Y$ depends on the current profile and for an isotropic current flow, $Y = 0.5$, while for a purely surface current $Y = 0$. The above integral is always logarithmically divergent, and because of the finite thickness of the strip, the double integral should exclude the domain where $|\bar{\rho}(\gamma) - \bar{\rho}(\beta)| < b$. A fair approximation is to use the expression

$$L \cong \frac{\mu_0}{4\pi} \int_0^\Theta \int_0^\Theta \frac{\Re\{\bar{\rho}'(\gamma)\bar{\rho}'^*(\beta)\}}{|\bar{\rho}(\gamma) - \bar{\rho}(\beta)| + b} d\gamma d\beta + \frac{\mu_0}{4\pi} Y \int_0^\Theta \rho(\gamma) d\gamma$$

(S24)

instead, which leads to numerically stable results. As a result of near-linear dependence of inductance to the displacement caused by magnetic stress

$$L = L_0 - \alpha x$$

(S25)

where $L_0 = 86.528$nH and $\alpha = \frac{1}{13.3}$fH/nm, the resonant frequency would also drift. It may be approximated as

$$\omega(x) = \frac{1}{\sqrt{L(x)C}} \cong \omega_0 \left(1 + \frac{\alpha}{2L_0} x\right)$$

(S26)

where $\omega_0 = 2\pi \times 2.4$GHz. Hence, we get the linear estimation

$$\frac{\partial \omega}{\partial x} \cong \frac{\alpha \omega_0}{2L_0} = 2\pi \times 1.051 \frac{\text{kHz}}{\text{nm}}$$

(S27)

This design has yet to be optimized with regard to the possible geometries and values of the inductance. The typical behavior does not essentially change much by changing the inductor's shape, such as $\rho(\theta) \sim \theta^m$.

With some modifications in the geometry, a great improvement may be achieved. However, changing material type with a difference stiffness does not help at all to change this value. But, it eventually enters the mechanical resonant frequency, in the sense that choosing a stiffer superconducting material such as p-doped Diamond, which is around 17 times more stiff than Aluminum, causes proportional increase in the resonant frequency, and hence decreasing the zero-point fluctuations $x_{zp}$. For this reason, the choice of Aluminum for the superconducting wire is probably the most suitable one.

For this purpose, we consider a few alternatives geometries, and go through identical calculations to observe the differences ($\Theta = 25\pi, a = 10\mu$m in all cases). The choice of $\Theta$ and $a$ together with the general prescribed forms of $\rho(\theta)$ ensures that all coils have the same area and number of turns, with the maximum radius $a\Theta = 0.785$mm. They are just different in their shape.

As opposed to the uniformly spaced spiral coil in (a), and as the coils get more compressed toward the periphery such as (b) and (c), the inductance value approaches a constant value [4] and $\partial \omega / \partial x$ gets saturated at the peak value of $\partial \omega / \partial x \cong 10$ kHz/nm.

However, if the coil is spinning more around the center such as (d), (e), and (f), the inductance value decreases while the $\partial \omega / \partial x$ attains much higher values. The reason is that the tail of the spiral is now less easy to deform under magnetic pinch pressure, since the tail of the spiral spring, which should be displaced more, is exposed to less magnetic field. As a result, and for the ultimately compacted coil, we obtain the value of $\partial \omega / \partial x = 0.19$ MHz/nm.

5

Purely mechanical property of spiral springs is a subject of deep study, and is published elsewhere [6-8]. But it is not difficult to evaluate the strain energy as follows.

The strain energy of a strained wire with cross section $S = hb$ and tensile strength $\sigma$ is given by [9]

$$W = \frac{1}{2}\sigma S \int_0^l \left(\frac{\partial u}{\partial x}\right)^2 dx = \frac{1}{2}\sigma S \int_0^\Theta \frac{[\Delta\rho'(\theta)]^2}{\rho(\theta)} d\theta$$

(S28)

where $u = \Delta\rho$ is the displacement along the wire, $dx = \rho(\theta)d\theta$ is the differential element of wire length, and $\Delta\rho'(\theta) = \partial\rho(\theta)/\partial\theta$ may be evaluated by direct differentiation of the first relation as

$$\Delta\rho'(\theta) = -\frac{\mu_0 I^2}{48\pi EM}\rho'(\theta)\rho^3(\theta)\int_0^\Theta \frac{\rho(\gamma)\Re\{[\bar\rho(\gamma) - \bar\rho(\theta)]e^{-i\gamma}\}}{|\bar\rho(\gamma) - \bar\rho(\theta)|^3} d\gamma$$

(S29)

Calculations give the following relationship for the strain energy versus current, which is almost quadratic in current, proportional to $I^4$.

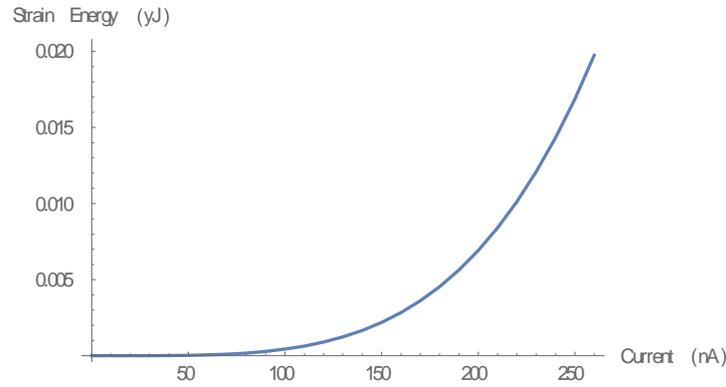

**Fig. S1.4.** Strain energy of a thin-wire suspended inductor versus current. The stored mechanical energy is roughly proportional to $I^4$.

The eigenfrequency of the string could be estimated by roughly the relationship

$$\cos(\beta_n l)\cosh(\beta_n l) + 1 = 0$$

(S30)

where

$$l = \int_0^\Theta \rho(\gamma)d\gamma$$

(S31)

is the total wire's length, and $\beta_n$ is the wavenumber of the $n$-th mode, given as

$$\beta_n = \left(\frac{\tau S}{EM}\right)^{\frac{1}{4}}\sqrt{\Omega_n}$$

(S32)

with $\Omega_n$ being the eigenfrequency, and $\tau$ the mass density of wire. To a good approximation [9], we have for the unstressed spring (with no current flowing)

$$\begin{aligned}\beta_1 l &= \lambda_1 = 1.8751 \\ \beta_2 l &= \lambda_2 = 4.6941 \\ \beta_3 l &= \lambda_3 = 7.8548 \\ \beta_n l &= \lambda_n = \left(n - \frac{1}{2}\right)\pi, \quad n \geq 4\end{aligned}$$

(S33)

For the configuration in the above, we thus get for the first four mechanical frequencies

| | |
|---|---|
| 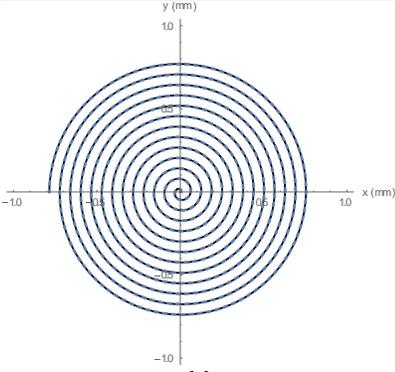<br>(a) | 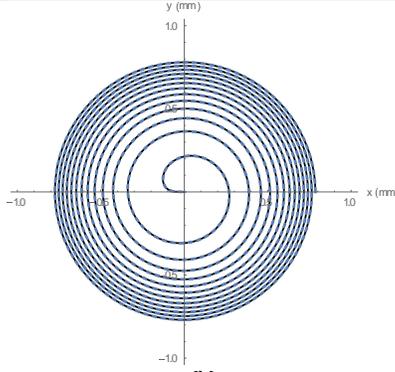<br>(b) |
| $\rho(\theta) = a\theta$<br>$L = 86.53\text{nH}, C = 50\text{fF}$<br>$\frac{\partial \omega}{\partial x} = 6.60 \frac{\text{kHz}}{\text{nm}}$<br>$g_0 = 50\text{mHz}$ | $\rho(\theta) = a\Theta \left(1 - \frac{\theta}{\Theta}\right)^{\frac{1}{3}}$<br>$L = 201.5\text{nH}, C = 30\text{fF}$<br>$\frac{\partial \omega}{\partial x} = 11.56 \frac{\text{kHz}}{\text{nm}}$<br>$g_0 = 89\text{mHz}$ |
| 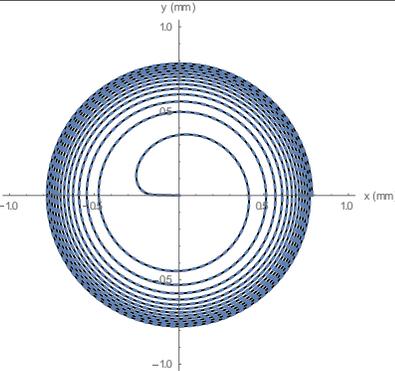<br>(c) | 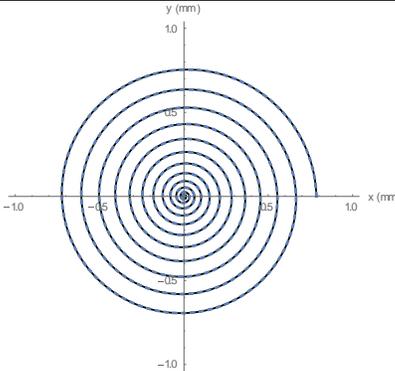<br>(d) |
| $\rho(\theta) = a\Theta \left(1 - \frac{\theta}{\Theta}\right)^{\frac{1}{5}}$<br>$L = 251.7\text{nH}, C = 30\text{fF}$<br>$\frac{\partial \omega}{\partial x} = 6.417 \frac{\text{kHz}}{\text{nm}}$<br>$g_0 = 49.3\text{mHz}$ | $\rho(\theta) = a\Theta \left(1 - \frac{\theta}{\Theta}\right)^{2}$<br>$L = 41.91\text{nH}, C = 50\text{fF}$<br>$\frac{\partial \omega}{\partial x} = 50.1 \frac{\text{kHz}}{\text{nm}}$<br>$g_0 = 0.386\text{Hz}$ |
| 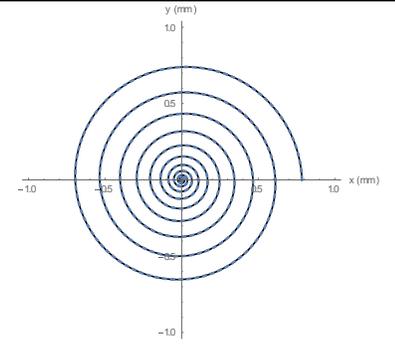<br>(e) | 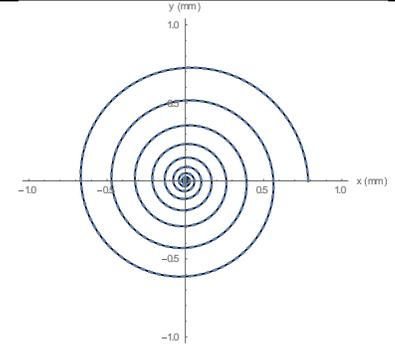<br>(f) |
| $\rho(\theta) = a\Theta \left(1 - \frac{\theta}{\Theta}\right)^{3}$<br>$L = 26.13\text{nH}, C = 50\text{fF}$<br>$\frac{\partial \omega}{\partial x} = 83.95 \frac{\text{kHz}}{\text{nm}}$<br>$g_0 = 0.645\text{Hz}$ | $\rho(\theta) = a\Theta \left(1 - \frac{\theta}{\Theta}\right)^{4}$<br>$L = 18.69\text{nH}, C = 20\text{fF}$<br>$\frac{\partial \omega}{\partial x} = 0.19 \frac{\text{MHz}}{\text{nm}}$<br>$g_0 = 1.46\text{Hz}$ |

**Table S1.1.** Dependence of the shape of suspended inductor on the interaction rate $g_0$.

$$\Omega_1 = 2\pi \times 5.7052 \text{ MHz}$$
$$\Omega_2 = 2\pi \times 35.74 \text{ MHz}$$
$$\Omega_3 = 2\pi \times 100.1 \text{ MHz}$$
$$\Omega_4 = 2\pi \times 196.2 \text{ MHz}$$
(S34)

The total wire's mass is given by $M = \tau l S$, which in this case is $M = 0.217\text{pg}$, and hence the motion mass is $m = 0.145\text{pg}$. Hence, we get the estimate for the zero-point fluctuations of the fundamental mode as

$$x_{zp} = \sqrt{\frac{\hbar}{2m\Omega_1}} = 8.0 \text{ fm}$$
(S35)

Hence, the single-photon coupling rate is given by

$$g_0 = \frac{\partial \omega}{\partial x} x_{zp} = 50.2 \text{mHz}$$
(S36)

This figure is too small, but may be significantly enhanced by proper design of the inductor. Estimations for $g_0$ is summarized in the table next page, and it has been demonstrated that $g_0$ may exceed 1Hz. COMSOL simulations should be done to obtain more accurate estimations of $g_0$ for spiral inductors.

### 1.1.3. Spiral Capacitor

As it was shown in Section 1.1.2, the out-plane deformation $\Delta\rho(\theta)$ due to a transverse electrostatic force was found, which would result in a net capacitance of

$$C = \epsilon_0 b B \int_0^\Theta \frac{\rho'(\beta)\rho^3(\beta)}{d - \Delta\rho(\beta)} d\beta$$
(S37)

Table 1 in the paper summarizes a variety of design spirals, cut through the same drum-capacitor structure while all the rest of the parameters are kept fixed. Here, for the first three rows, it is essentially the number of spiral turns which is varied and the strip width and spiral spacing are changed accordingly. Calculations are done using (S37) and $\Delta\rho(\theta)$ profile fed from COMSOL calculations. However, as it is being shown in the below, $\Delta\rho(\theta)$ could be still estimated from (S17) but that would cause a large underestimation of $g_0$, implying that just a few number of turns should be quite sufficient to obtain a significantly large enhancement of $g_0$.

Anyhow, in order to observe what the thin-wire model would yield, we may take the electrode spacing as $d = 50\text{nm}$, and then a 50fF capacitor will need a total strip length of 14.12mm. This will fit into a circle of radius 13.4µm, using a uniform spiral shape of $N = 335$ turns. This number of turns is unrealistic and unnecessarily large as a much smaller number of turns should in principle according to the Table 1 already result in a sufficiently large $g_0$. For this reason, we conclude that while the thin-wire model is conceptually correct within the order of magnitude, its results are not useful. The main reason for the discrepancy is that the 1D cantilever approximation is too wrong for a spiral structure, and overestimates the true fundamental mechanical frequency by many orders of magnitude.

As it has been argued in the article, the cantilever approximation cannot be used for modeling spiral structures and is highly inappropriate. Therefore, the deformation profiles are obtained from COMSOL calculations and fed into the relationship (S37) to obtain much more accurate results. This has been illustrated in the next two figures.

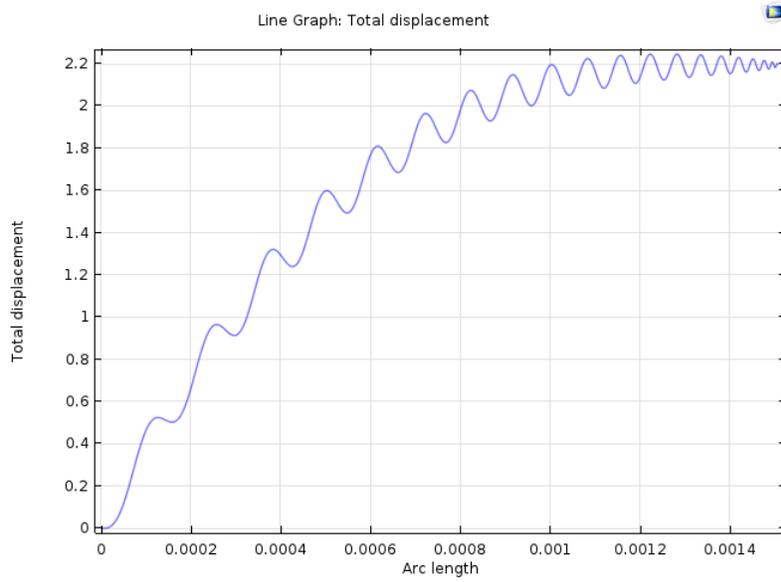

**Fig. S1.5.** Normalized deformation profile along the spiral strip length for $N = 20$.

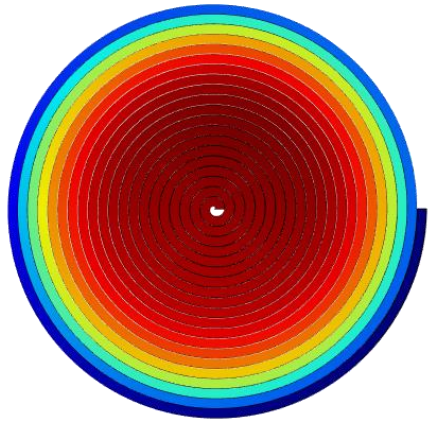

**Fig. S1.6.** Fundamental mechanical mode of the spiral capacitor for $N = 20$.

The dependence of capacitance on displacement from the thin-wire approximation combined with COMSOL input is found to be given as below shown in the next Figure.

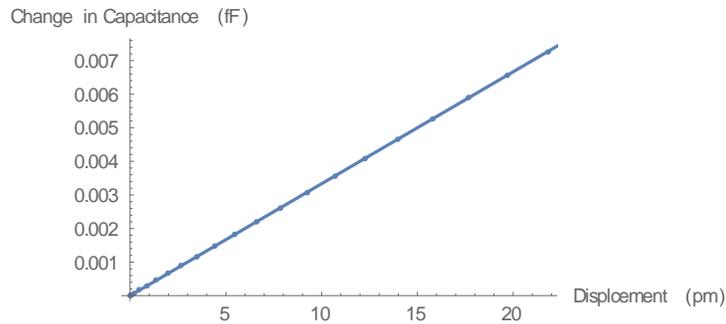

**Fig. S1.7.** Change in spiral capacitance versus out-plane displacement due to the electrostatic attraction of the electrodes.

The ultimate general scaling law at constant capacitance would be looking something like this

$$g_0 \sim \frac{1}{b\sqrt{dhL_0}}$$

(S38)

9

which is surprisingly more or less independent of the gap $t$. A further inspection using COMSOL simulations, and noting $r = N(b+t) \sim Nb$ with $b \gg t$ yields a slightly different result as

$$g_0 \sim \sqrt{\frac{N}{dhL_0}}$$

(S39)

leading to the general result that the single-photon interaction rate increases with the square root of the number of turns $g_0 \sim \sqrt{N}$.

Finally, for the same reason that the 1D thin-wire cantilever is highly inappropriate for modeling of spiral capacitors, it should not be considered to be reliable for modeling of spiral inductors, since fundamental mechanical frequencies would be highly overestimated. It is expectable that the estimated $g_0$ values in Table 2.1 for suspended spiral inductors have been underestimated roughly by an order of magnitude or even higher. This fact raises hopes for a wider range of applications and usefulness of such electromechanical elements.

Therefore, again COMSOL simulations should be trustable for obtaining mechanical resonant frequency as well as deformation profiles. A full integrated electromechanical COMSOL simulation is not necessary since the electrostatic or magnetostatic analysis could be more or less combined in Mathematica to the COMSOL output profiles. While this significantly increases the simulation speed, it would complicate the steps required to carry out a full scale simulation.

## 2. SIMULATION

Simulations are done in COMSOL 5.2 Multiphysics software on the LPQM simulation server. Each run typically takes a few of minutes to an hour and it is fairly easy to test various structures with different parameters. For this purpose, the Solid Mechanics module is fine enough and a few setting parameters can be defined, similar to Table S1.1 which define the shape and geometry of the spiral. This is shown in Fig. S2.1.

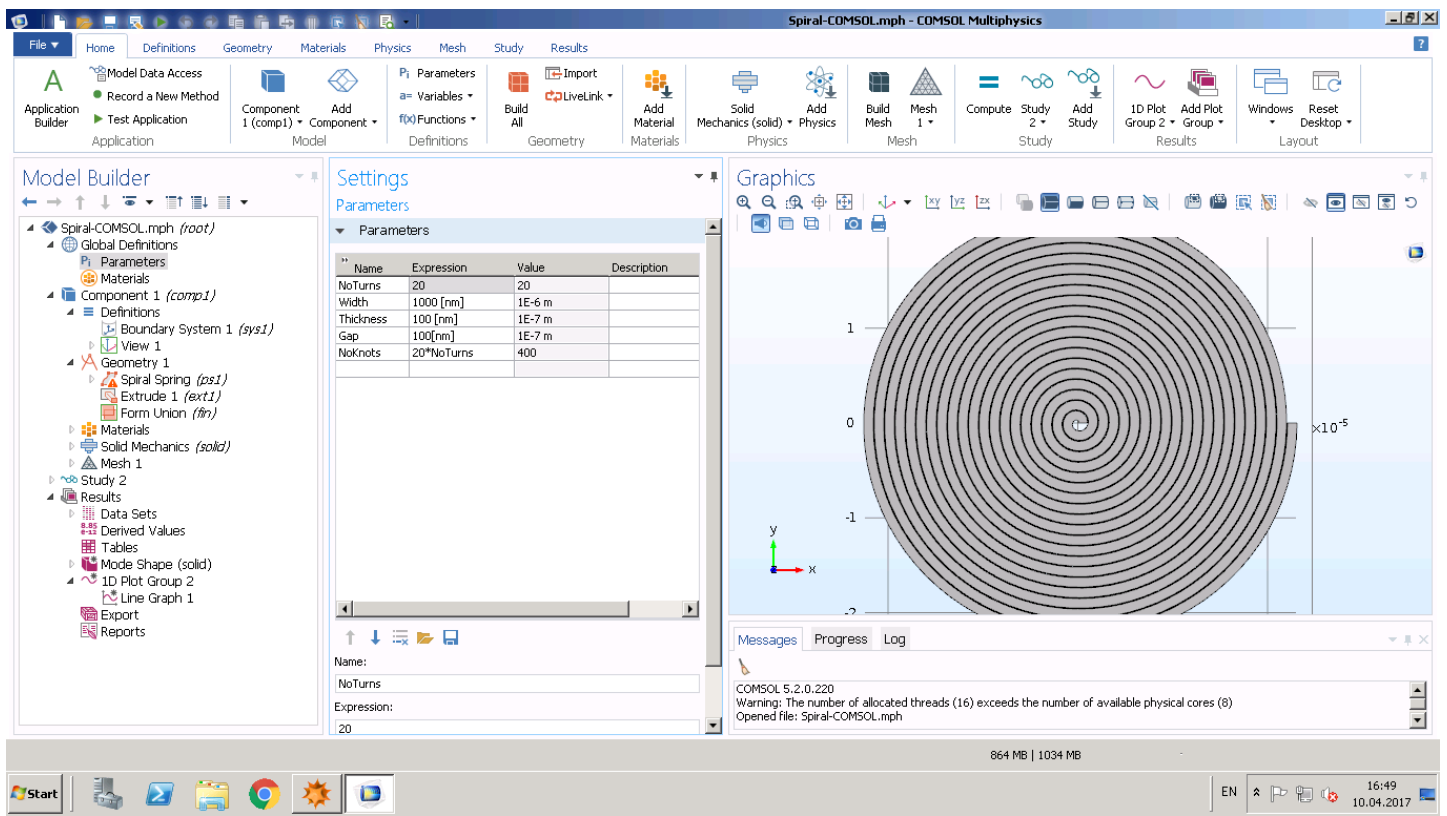

**Fig. S2.1.** Initial definition of parameters.

The spiral shape can be defined by geometric functions in parametric polar coordinates, which is shown in Fig. S2.2. Then it needs to be extruded in the third dimension to make it of finite thickness. Afterwards, the outer boundary should be fixed to

allow a cantilever-like oscillation along the spiral. This is done by fixing the corresponding boundary, which is shown in Fig. S2.4.

Afterwards, it would be possible to obtain the eigenmodes and profiles in a straightforward manner. The degree of coarseness of meshes is noticed to be almost regardless of the quality of results, so that the choice of 'Fine' is good enough. For this example, the first and second eigenmodes are at the frequencies of 1.6kHz and 3kHz. The second eigenmode is shown in Fig. S2.5 and the corresponding deformation profile is plotted in Fig. S2.6. Hence, the fundamental mechanical mode is isolated to a good accuracy from the next modes.

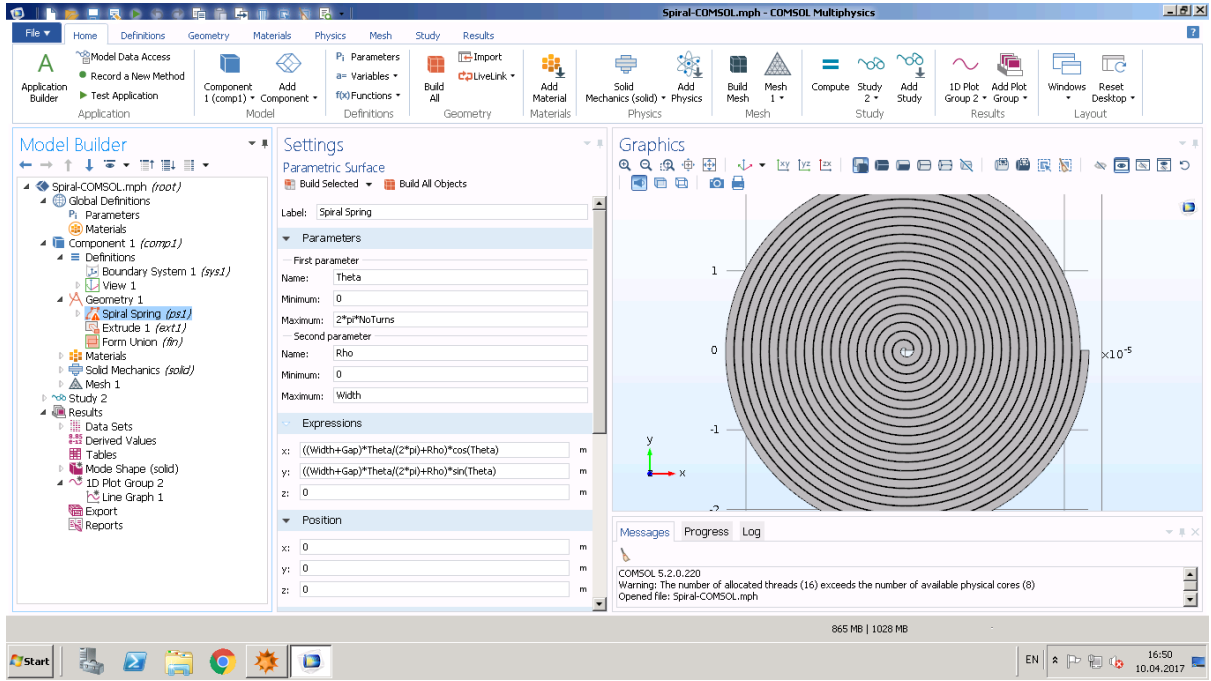

**Fig. S2.2.** Definition of parametric surface for construction of spiral with $N = 20$.

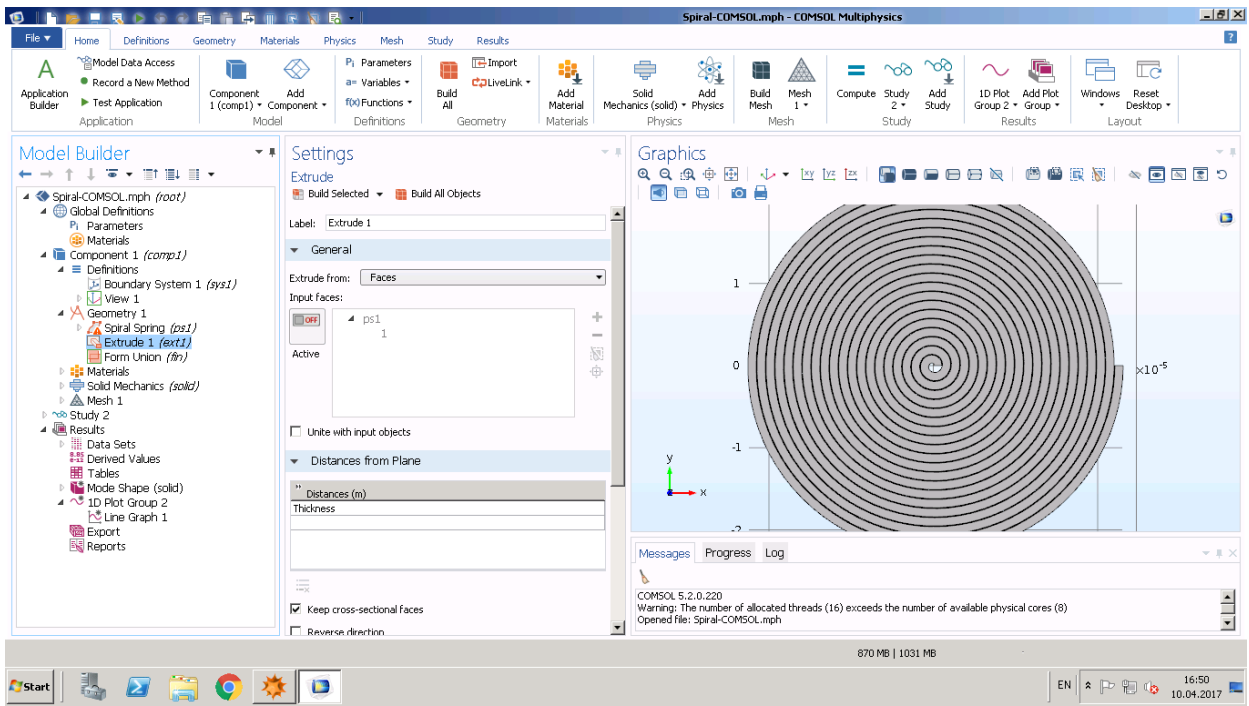

**Fig. S2.3.** Extrusion of the spiral to give it a finite thickness with $N = 20$.

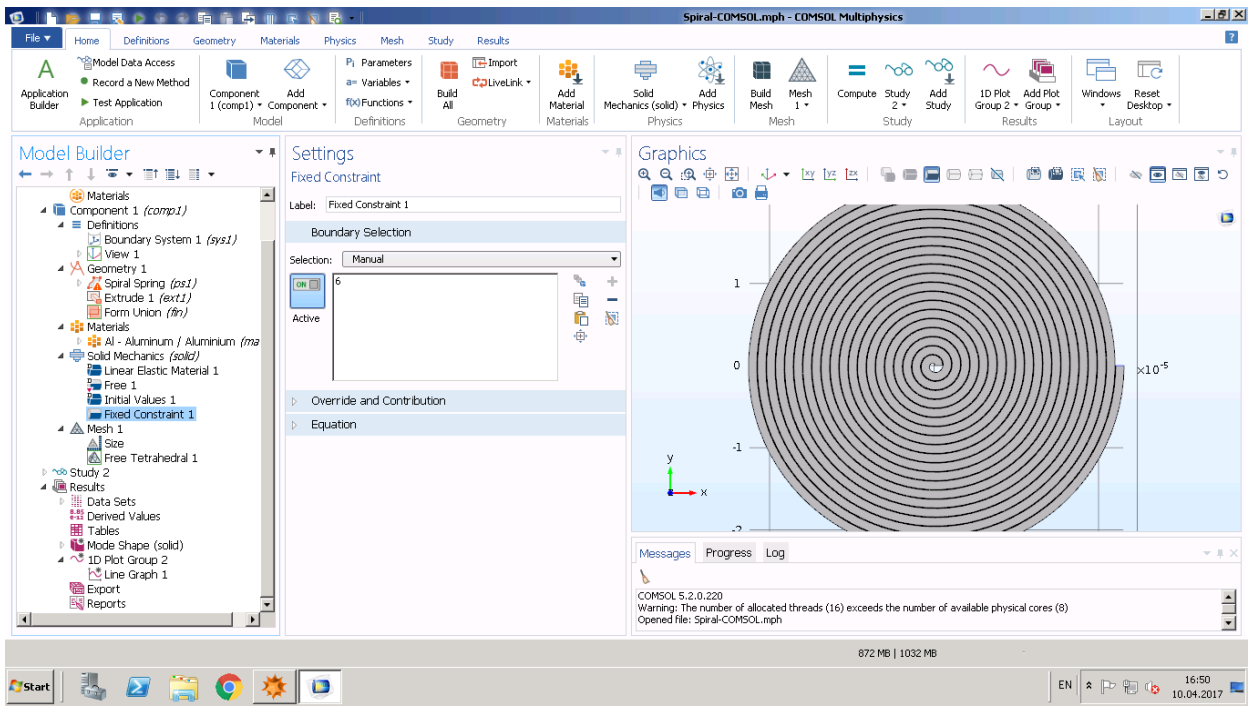

**Fig. S2.4.** Fixing the boundary with $N = 20$.

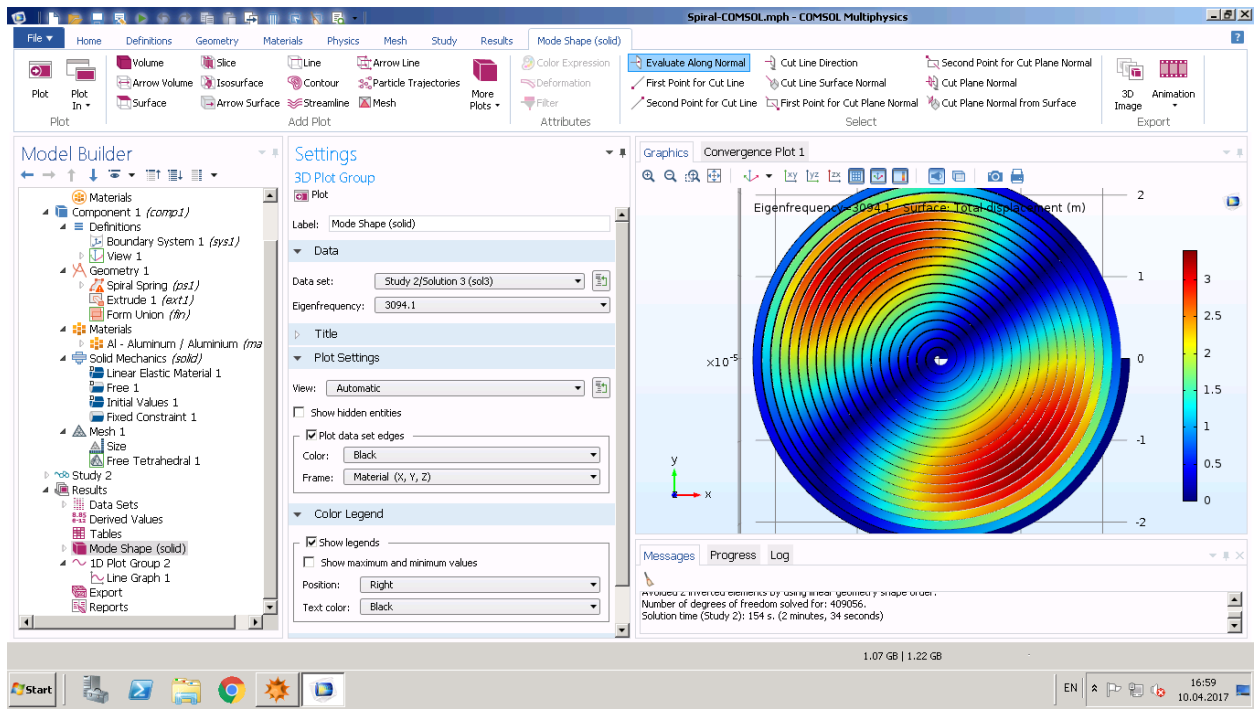

**Fig. S2.5.** Second mechanical eigenmode with $N = 20$.
That of the fundamental mode is already shown in Fig. 2.5.

Should anyone want to check $N = 5$ instead of $N = 20$, it is only sufficient to go to the initial step as in Fig. S2.1, set the desired number of turns and strip width and spacing, which we set according to the Table S1.1, and rerun the module. Doing 'Build All' immediately results in the updated structure as in Fig. S2.7. The resulting fundamental mode looks like as shown in Fig. S2.8.

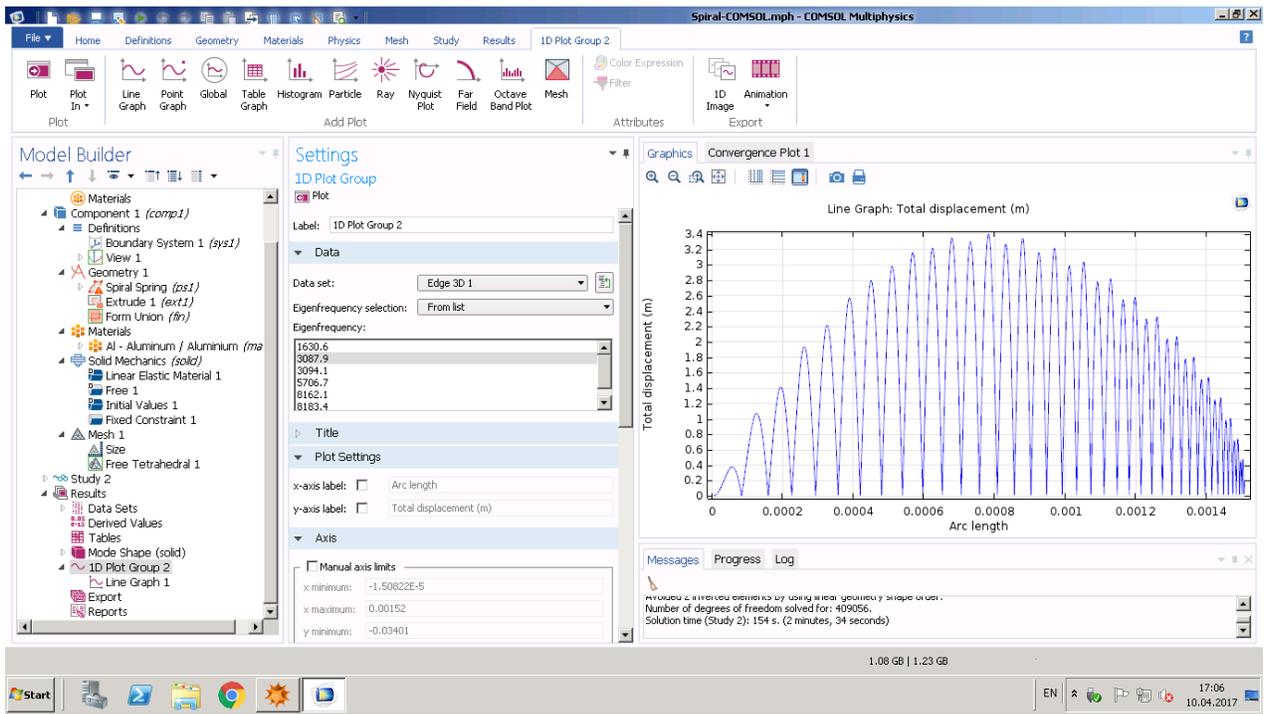

**Fig. S2.6.** Deformation profile of the second mechanical eigenmode with $N = 20$.
That of the fundamental mode is already shown in Fig. S1.5.

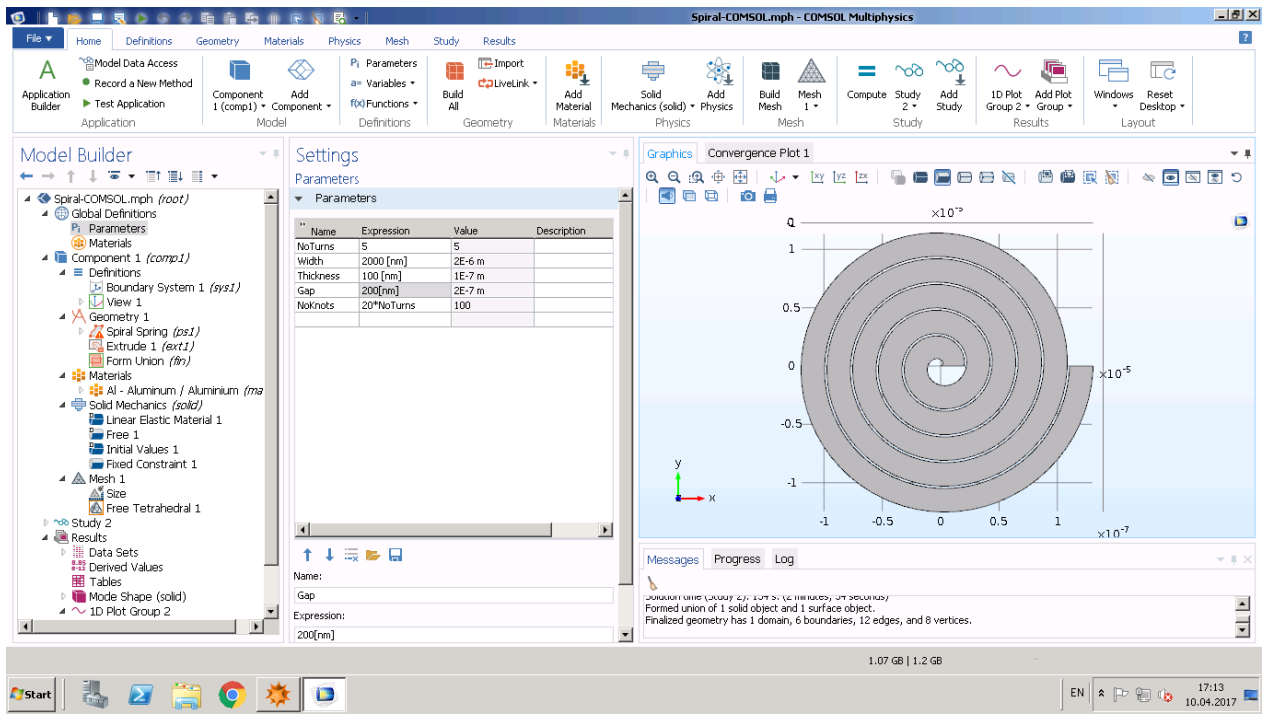

**Fig. S2.7.** Updated spiral with $N = 5$ as the number of turns.

## 3. FABRICATION

We have not tried fabrication of the suspended inductor. This is both much easier in fabrication and much larger in size compared to spiral capacitor. For a suspended spiral capacitor, a strip of thickness $h$ and width $b$ is assumed to form a spiral shape with gap $t$. Since the total spiral is equipotential, the smallness of gap is irrelevant. So, one may reduce the gap and width to obtain many turns of spiral and thus total wire length $l$. The capacitor from the top looks like below as shown in Fig. S3.1.

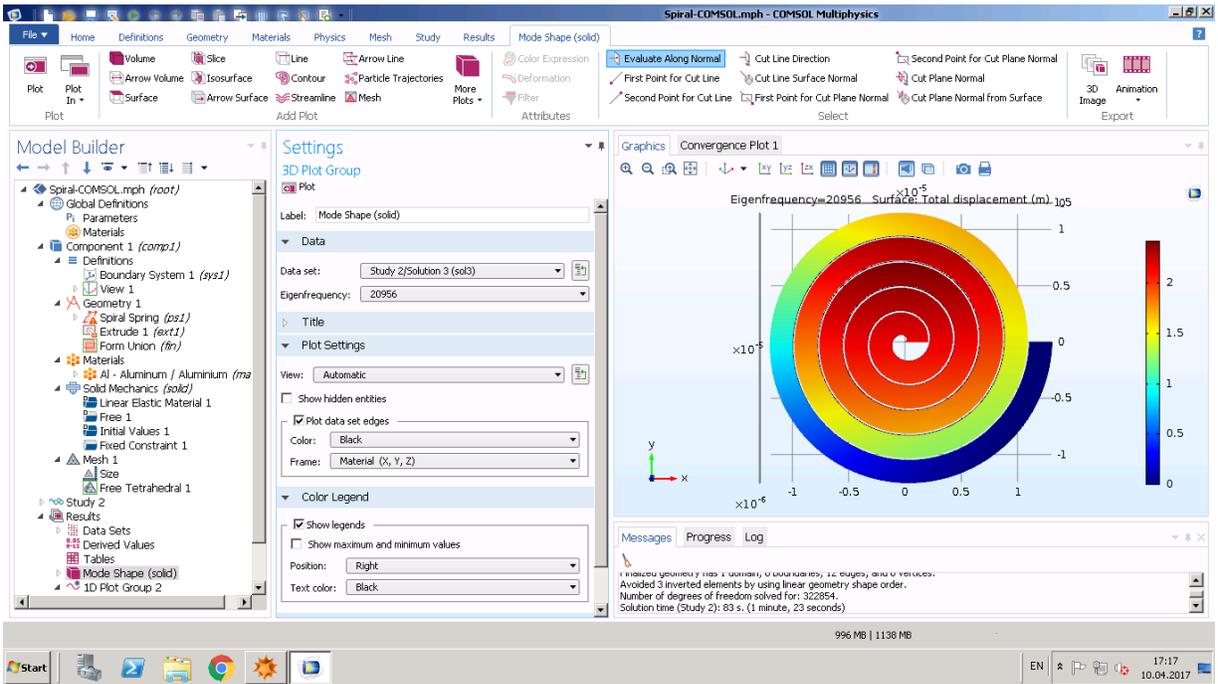

**Fig. S2.8.** Fundamental mode of the spiral with $N = 5$.

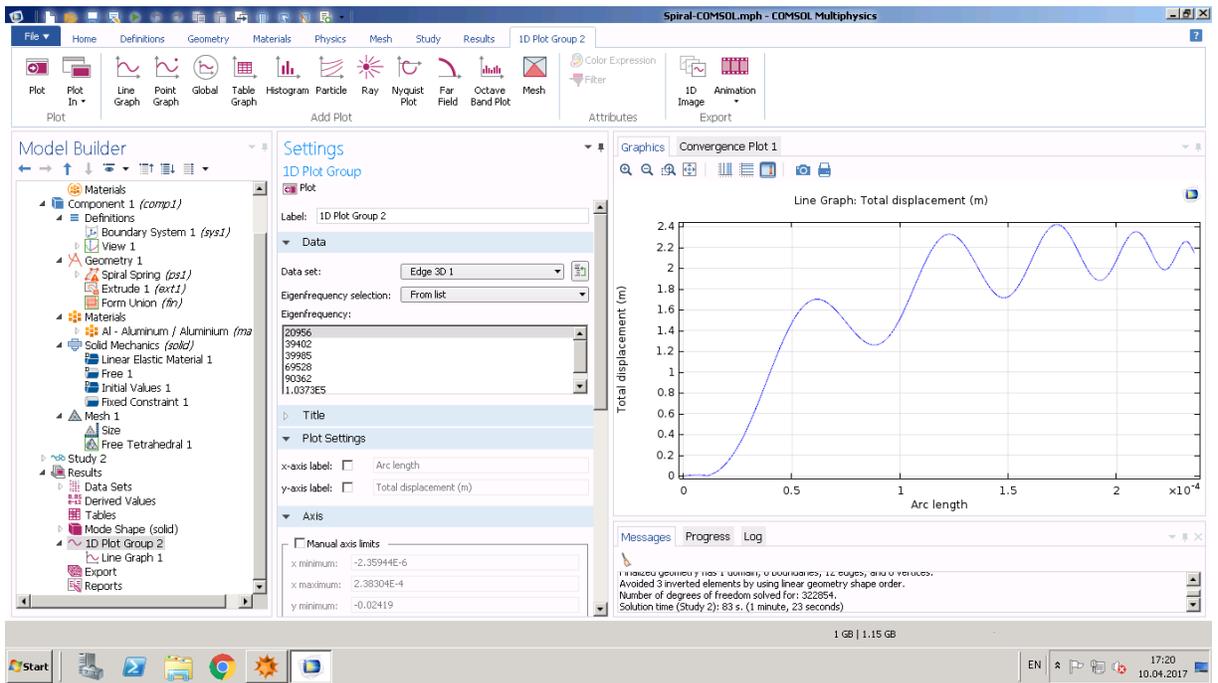

**Fig. S2.9.** Deformation profile of the fundamental mode of the spiral with $N = 5$.

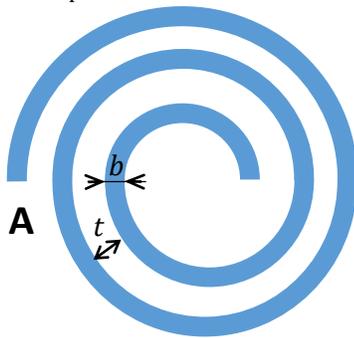

**Fig. S3.1.** Top view of the suggested spiral drum capacitor.

14

What we have actually done is to carve an ultrathin spiral gap using FIB tool unto the micro-drum capacitor prior to the undercutting and release. This not only speeds up prototyping but also avoids many extra steps needed in repeating the previous fabrication steps, which should have been otherwise redone using updated masks and process.

The present micro-drum capacitor is having a diameter of 24μm, so that the minimum feature size is too large to require any high resolution lithography such as Electron Beam Lithography (EBL). For the parameters chosen, the spiral gap is only 200nm, and features as small as 20nm could be carved out well deep into the 100nm thick Aluminum. Hence, the spiral could be made by FIB only in a single additional step and done maskless. The resulting Al strip width would be 2μm with $N = 5$. Fig. S3.2. illustrates the false-colored SEM photo of the unpatterned micro-drum capacitor, and the intended spiral cut.

**Fig. S3.2.** Normally used drum capacitor with flat circular membrane top electrode [10] and suggested spiral top electrode.

**Fig. S3.3.** Fabrication process flow [10] for Fig. S3.1 is the same as Fig. S3.2, except that the lithography pattern of the top electrode is more detailed.

It is also quite possible to redesign all masks and remake them using EBL, so that this additional FIB step could be removed. However, UV lithography would not be possible anymore and the compatibility of E-Beam photoresists with the present process has yet to be investigated. Therefore, unless a rigorous process has to be developed from the scratch, probably the most straightforward way to fabricate a spiral capacitor at LPQM is to use the present micro-drum capacitors before undercut and release, take them to the FIB, do the patterning, and then carry out the structure release and undercut at last.

The accuracy of UV lithography employed in the previous steps is good enough to support fabrication of suspended inductors if necessary. However, the corresponding Aluminum thickness probably needs to be much more than 100nm to provide sufficient mechanical strength to support the structure weight. These issues need further investigation. Should masks need redesign to be fabricated with EBL, then there exist macros such as the one in the Box S3.1 of the next page which could easily put a spiral to the mask design with desired shape parameters. The typical output of Layout Editor due to this macro is shown in Fig. S3.4.

```
  1 #!/usr/bin/layout
  2 #name=Multiturn Spiral
  3 #help= mai 30 2008 Serge Charlebois
  4 /*      Creates a new cell and draws a multiturn spiral in it.
  5         The spiral is centered on (0,0).
  6         Requests from the user the number of turns, trace width and spacing and the termination type.
  7         The cell name is forced to "Spiral#" where # is the number of turns.
  8 */
  9
 10 int main(){
 11 int i, nt, cap;
 12 point origin, start, end;
 13 double step, gap, width, microns;
 14 string cellname;
 15
 16 microns = 1/layout->drawing->userunits;
 17 origin.setX(0*microns);
 18 origin.setY(0*microns);
 19
 20 nt = layout->getInteger("User entry","Number of turns for the spiral:");
 21 if (nt==0){
 22         layout->showMessage("Error message","The spiral needs at least 1 turn. Macro aborted.");
 23         return 0;
 24 }
 25 width = microns * layout->getDouble("User entry","Width of the traces (userunits):");
 26 gap = microns * layout->getDouble("User entry","Distance between traces (userunits):");
 27 cap = layout->getInteger("User entry","Termination type: 0 for none, 1 for circle and 2 for square");
 28
 29 step = width + gap;
 30 cellname.setNum(nt);
 31 cellname = "Spirale"+cellname;
 32
 33 // add new empty cell
 34 cellList *cl=layout->drawing->addCell();
 35 // set new cellname
 36 cl->thisCell->cellName = cellname;
 37 // set new cell to current cell
 38 layout->drawing->setCell(cellname);
 39
 40
 41 // loop on all sprial turns
 42 i=0;
 43 while (i<nt){
 44
 45 // calculate start point of one turn
 46 start.set(origin.y(),origin.x() + i*step);
 47 end.set(origin.y(),origin.x() + (i+1)*step);
 48 // calculate end point of one turn
 49
 50 // create one spiral "turn"
 51 layout->drawing->point(origin);
 52 layout->drawing->point(start);
 53 layout->drawing->point(end);
 54 layout->drawing->spiral();
 55 layout->drawing->currentCell->firstElement->thisElement->setWidth(width);
 56 layout->drawing->currentCell->firstElement->thisElement->setCap(cap);
 57
 58 i++;
 59 }
 60
 61 // refresh the screen by adjusting the display scale
 62 layout->drawing->scaleFull();
 63
 64 return 0;
 65 }
```

**Box 3S.1.** Layout Editor's macro code to generate a spiral pattern.
Similar codes with other softwares could be used to produce the desired EBL mask.
Source: http://www.layouteditor.net/wiki/SpiraleMacro

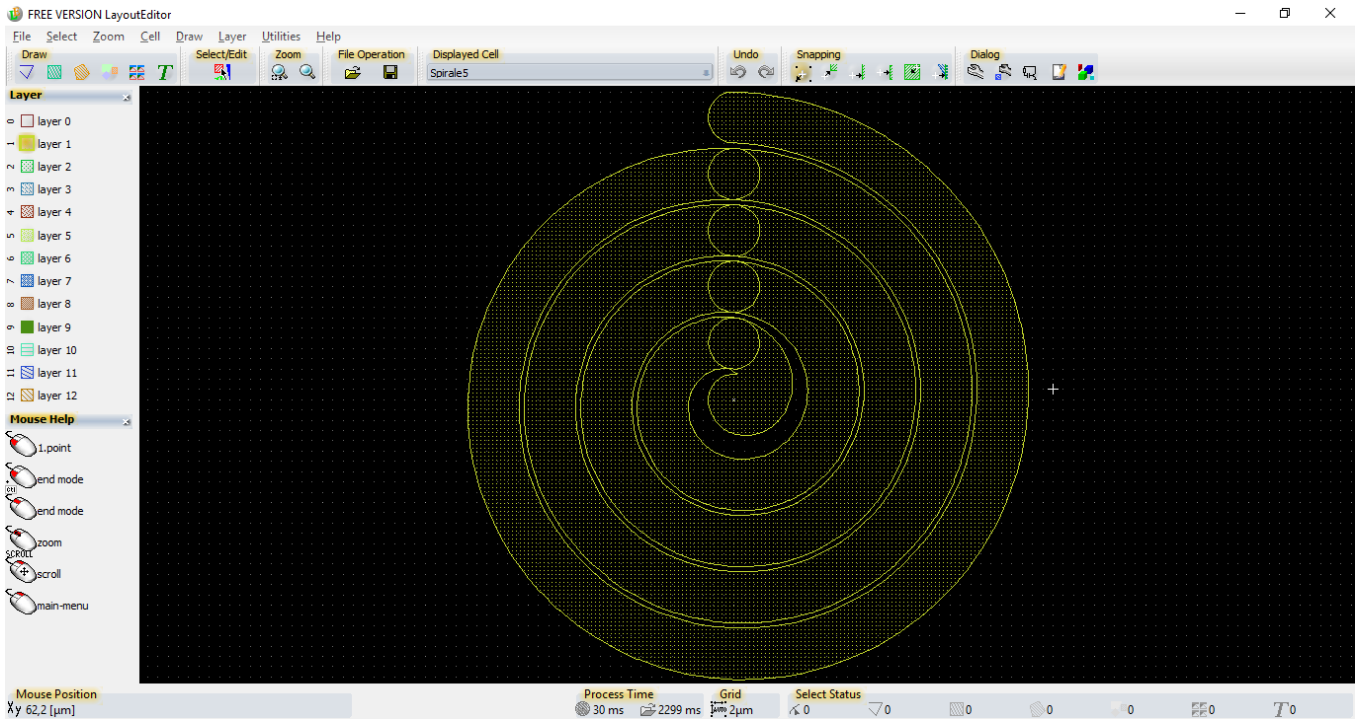

**Fig. S3.4.** Typical output of the Layout Editor macro in Box S3.1.

### 3.1. Process Flow Used at CMi

In what follows, the process flow is presented which was submitted to the EPFL's CMi and was approved after numerous edits and improvements.

| Technologies used |
|---|
| DC sputtering (Aluminum, Amorphous Silicon), UV Photolithography (MLA 150), $XeF_2$ etching, FIB, Dicing, SEM, STS/Cl Dry Etch, $O_2$ Asher + Wet Remover 1165, AMS or STS/HBr Dry Etch |
| **Substrate Type** |
| High Resistivity Undoped Silicon Wafer <100>, Ø100mm, 500um thick, Single Side polished |

**Table S3.1.** Process requisites and initial materials.

**Table S3.2.** Process steps with Focus Ion Beam. Fabrication of the capacitor could be possibly simplified by removing step 08 and merging into step 04. However, UV lithography could no longer be used for this step and it should be replaced by EBL.

| Step | Process description | Cross-section after process |
|---|---|---|
| 01 | Substrate:<br>High Resistivity Undoped Silicon Wafer ($\rho$ > 5000 $\Omega$cm)<br>RCA cleaning | |
| 02 | 100nm $SiO_2$<br>Dry Oxidation | |
| 03 | 100nm Al Deposition via DC Sputtering | |

17

| | | |
|---|---|---|
| **04** | UV Photoresist Coating & Lithography (MLA 150) | 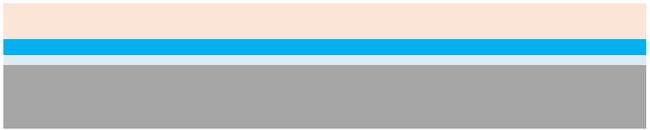 |
| | Photoresist Development | 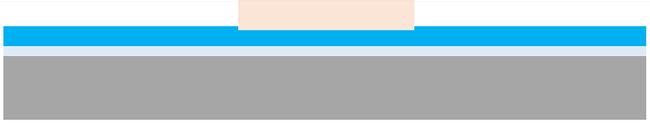 |
| | STS Dry Etch of Aluminum with Chlorine Chemistry | 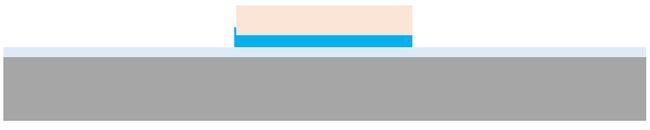 |
| | Photoresist Strip via O$_2$ Asher + Wet Remover 1165 | 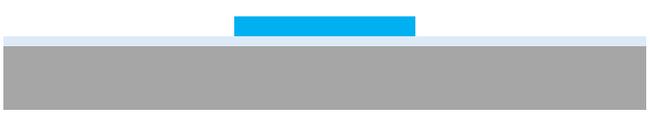 |
| **05** | 100nm Amorphous Silicon Sacrificial Layer via DC Sputtering | 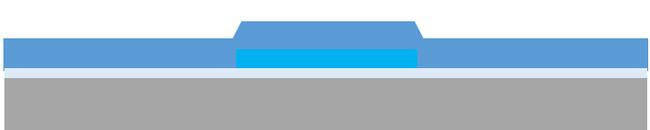 |
| **06** | 1-2 µm Photoresist thick coating | 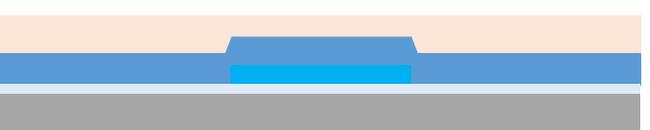 |
| | UV Photolithography and Development (MLA 150) | 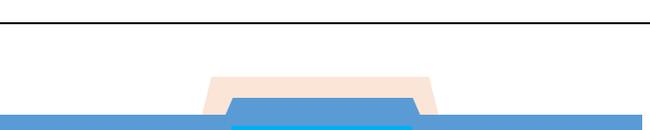 |
| | AMS Dry Etch of Aluminum or STS with HBr Chemistry | 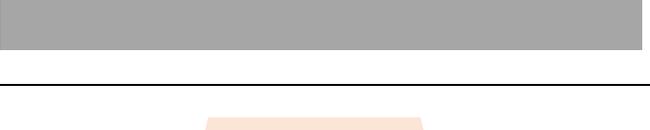 |
| | Photoresist Strip via O$_2$ Asher + Wet Remover 1165 | 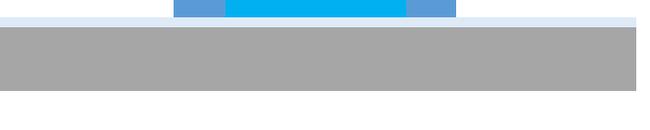 |

| | | |
|---|---|---|
| 07 | 100nm Al Deposition via DC Sputtering | 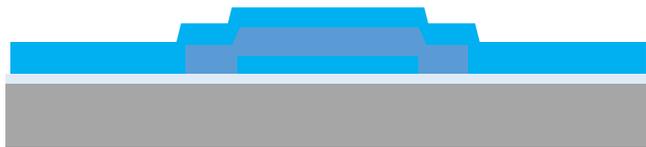 |
| 08 | FIB Aluminum Etch (High Resolution; Small Exposure Area ~ 1000μm$^2$) 30kV, 0.3nA | 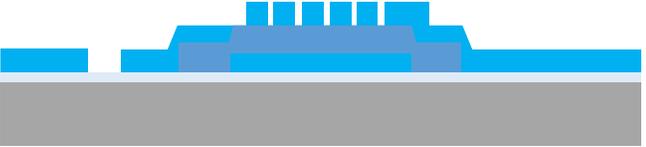 |
| 09 | XeF$_2$ Etch of Amorphous Silicon Sacrifical Layer | 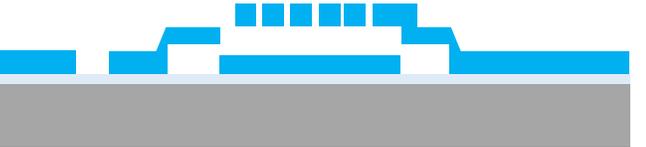 |

**3.2. Fabrication Results**

Two runs of the FIB were done. The first run (in November 2016) deposited too much charge unto the surface which resulted in loss of focus. Hence, the FIB should have re-written quite a few times. As a result of too much heat imposed on the very thin Aluminum layer, which is only 100nm thick, parts of the connecting wire was molten and evaporated away. The pattern has to be prepared in Black-and-White pixel image, with white pixels representing points to be scanned by the focused ion-beam. This is illustrated in Fig. S3.5. The user should not be worried much about the scaling, since it is possible to fix the physical size on the available software before patterning.

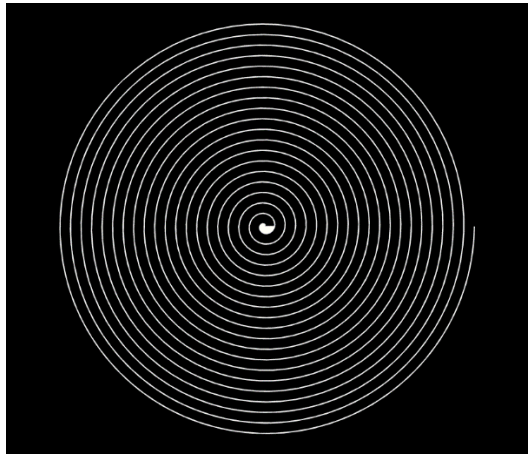

**Fig. S3.5.** Image supplied to the FIB machine for patterning.

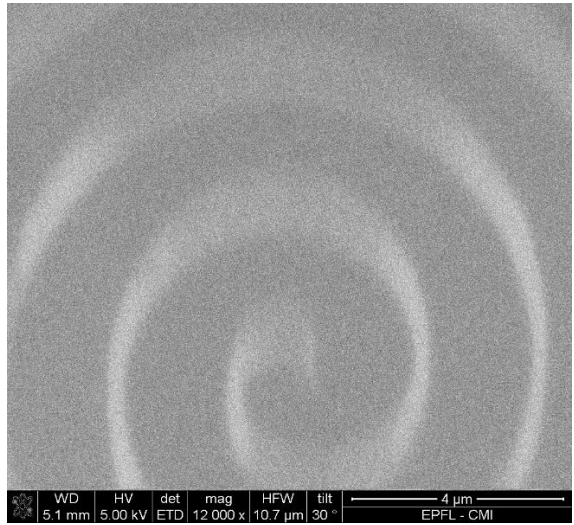

**Fig. S3.6.** Initial loss of focus in the fabrication of first sample.
The image is taken with the SEM tool of the FIB machine.

One of features of the FIB machine is the possibility of taking SEM and Ion-beam images at once, and this could be done while patterning as well. Figs. S3.8 and S3.9 respectively illustrate the fabricated spiral under the electron-beam and ion-beam. Only one of these two images could be taken at the normal angle and at the same time, since the electron and ion guns make an angle of roughly 57 degrees. Since the substrate is non-conducting, the FIB patterning should be done in a single run, otherwise too much charge will be deposited unto the surface which results in loss of a sharp image and accuracy.

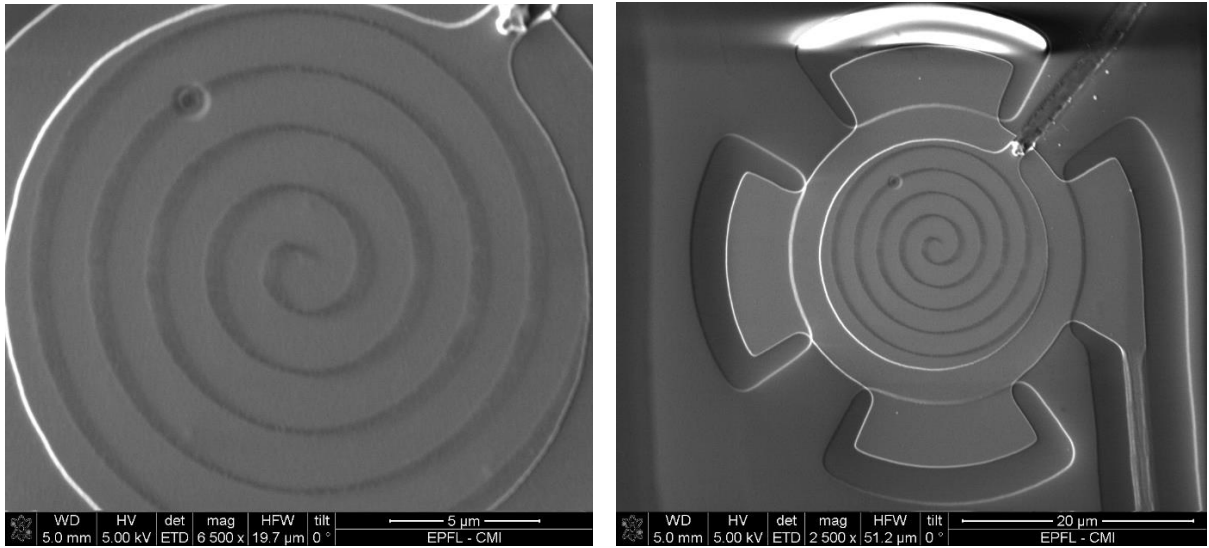

**Fig. S3.7.** Second illumination by FIB in the first sample. Too much dosage caused blow up of one connecting wires, and electrical disconnection of the capacitor.

A second sample was therefore fabricated in early February 2017, since the first sample had been already electrically disconnected and furthermore it had been somehow lost at the clean room. This time, the sample was patterned much better and more accurately compared to the first one, and it remained electrically connected.

While the first sample was never taken to the undercutting process of XeF$_2$ etch, the second sample was. It can be seen that it has survived the undercut very well. We could carry it around afterwards quite safely to the SEM zone and take a few additional SEM photos from the normal and oblique incidence. The SEM photos reveal that the spiral capacitor structure with *N*=5 has actually survived the undercut very well. It should be added that since the width of gaps is less than 200nm, the spiral is totally invisible under an optical microscope. Only SEM could reveal the spiral feature, and then it is very visible and quite obvious.

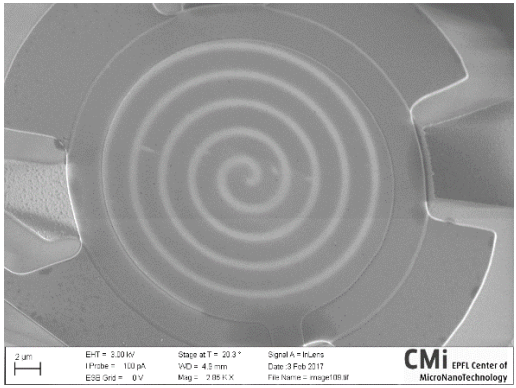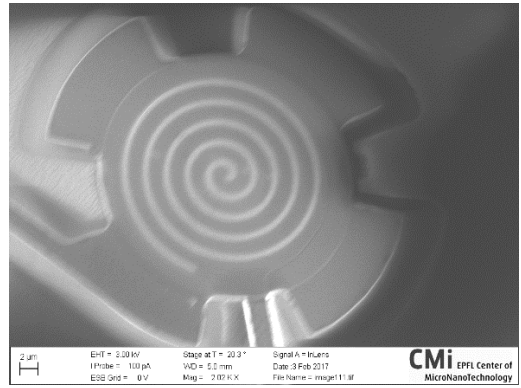

**Fig. S3.8.** Oblique SEM after XeF$_2$ undercut and structure release.

 
\end{document}